\documentclass[pre,twocolumn,showpacs,preprintnumbers,amsmath,amssymb]{revtex4}
\usepackage{simplewick}

\usepackage{graphicx}
\usepackage{dcolumn}
\usepackage{bm}
\newcommand \be{\begin{eqnarray}}
\newcommand \ee{\end{eqnarray}}
\newcommand \ba{\begin{align}}
\newcommand \eea{\end{align}}

\begin{document}

\title
{Phase-field theory of brine entrapment in sea ice: Short-time frozen microstructures}
\author
{
Silke Thoms$^1$, Bernd Kutschan$^2$, Klaus Morawetz$^{2,3,4}$ 
}

\affiliation{%
$^1$Alfred Wegener Institut, Am Handelshafen 12, D-27570 Bremerhaven, Germany}
\affiliation{%
$^2$ M\"unster University of Applied Science, Stegerwaldstrasse 39, 48565 Steinfurt, Germany}
\affiliation{$^3$International Institute of Physics (IIP)
Federal University of Rio Grande do Norte
Av. Odilon Gomes de Lima 1722, 59078-400 Natal, Brazil}
\affiliation{$^4$Max-Planck-Institute for the Physics of Complex Systems, 01187 Dresden, Germany}

\begin{abstract}

We analyze the early phase of brine entrapment in sea ice, using a phase field model. 
This model for a first-order phase transition couples non-conserved order parameter kinetics 
to salt diffusion. The evolution equations are derived from a Landau-Ginzburg order parameter
gradient dynamics together with salinity conservation. The numerical solution of model equations 
by an exponential time differencing scheme describes the time evolution of phase separation 
between liquid water with high salinity and the ice phase with low salinity. The numerical solution
in one and two dimensions indicates the formation of one dominant wavelength which sets the length
scale of short-time frozen structures. A stability analysis provides the phase diagram in terms of
two Landau parameters. It is distinguished an uniform ice phase, a homogeneous liquid saline water
solution and a phase where solidification structures can be formed. 
The Landau parameters are extracted from the supercooling and superheating as well as the
freezing point temperature of water. With the help of realistic parameters the distribution of
brine inclusions is calculated and found in agreement with the measured samples.
The size of the ice domains separating regions of concentrated seawater depends on salinity and
temperature and corresponds to the size of sea ice platelets obtained from a morphological
stability analysis for the solidification of salt water.

\end{abstract}
\maketitle

\section{Introduction}

The formation of sea ice plays an important role in earth's climate because it determines 
the large-scale heat and mass transport delivered to the surface of the polar oceans.
Dependent on the season, sea ice covers an area of around $(4.3-15.7)\times 10^6 km^2$ in 
the Arctic and between $(3.6-18.8)\times 10^6 km^2$ in the Antarctic \cite{Arr,DieHel10}. 
Thus, at its maximum extent, sea ice covers about $10\%$ of the ocean's surface area.

A large number of brine inclusions emerge in sea ice when ice is formed.
During the freezing process, solutes in the seawater are excluded from the ice matrix 
and segregate into brine droplets or in intercrystalline brine layers and at grain 
boundaries \cite{Weeks86,Ei}. The size and distribution of brine inclusions 
depend on the thermal history and growth conditions.

Despite the extreme growth conditions of low temperatures and high salinities, sea ice offers 
a favorable environment for a variety of organisms, particularly bacteria and microalgae, which 
live and thrive within the brine inclusions. Brine entrapment in sea ice is a crucial 
element in the polar ecosystem because it is an important habitat for a variety 
of CO$_2$-binding microalgae. A high lasting biomass yield 
of $10-1000$ mg Chl m$^{-3}$ \cite{Arr01,Liz01,Thomas02} has been measured within sea ice, 
whereas in the water under the ice chlorophyll is usually below $0.1$ mg m$^{-3}$ \cite{Liz01}. 
Thus, due to the ice-algal growth a concentrated food source is delivered to the low-productivity
ice-covered sea. In the months of ice melting, ice algae can provide a starter community for
phytoplankton growth by seeding the water column \cite{Liz01}.
To unravel the interplay between the microorganisms and their natural habitat the knowledge of 
the pore space connectivity and the conditions under which different inclusion morphologies 
are formed is mandatory.

Understanding the solidification of salt water is a long-standing interest. 
Already {\sc Quincke} considers ice polluted with any salt as ''liquid jelly'' with walls 
and cells {\cite{Qui2}}. The freezing process of salt water is but one example of the 
solidification of binary alloys. The thermodynamics and solidification theory of alloys 
(see e.g. \cite{Till53,Chal64}) have been used in a variety of forms to study
solidification systems in the context of metallurgical and geological applications. 
The two-phase region comprising of essentially pure ice crystals bathed in water enriched in the
rejected salt is known in the context of binary alloys as a mushy layer. 
The theory of mushy layers has been applied to explain the bulk features of sea ice. 
Rather than modelling the small-scale spatial variability, the theory of mushy layers provides
predictions of the properties of the mushy layer averaged over the microscale. The equations 
describing the conservation of mass, heat and salt predict the dynamics of growing sea ice in 
terms of local mean variables such as temperature and the concentration of the interstitial 
liquid. Darcy's equation is used to calculate the local mean brine flux in the rigid matrix 
of ice which is described in terms of the local mean porosity \cite{WW97,Felt06}.

From a small-scale perspective  {\sc Golden} {\it et al.} \cite{Gol, Gol1, Gol2} discussed 
the role of percolation transitions in brine trapping during the solidification of seawater. 
The distribution of brine entrapments and its impact on fluid- and nutrient transport is reported 
in \cite{Gol2}. As key variable for all transport processes the permeability was introduced. 
Permeability depends on porosity, as well as on microstructural characteristics of brine inclusions.
Porosity, defined as volume fraction of brine inclusions in total ice volume, is a determining
feature for biological activity within ice. The microstructure near the ice bottom is particularly
important for algae growth. Highest cell abundances occur in this region, due to the higher porosity
and to the constant flushing with nutrient-rich seawater \cite{Ackl94,Wer07}.

Mushy layers are formed due to a instability of the solidifying interface and are ubiquitous 
during the solidification of alloys. The theory of morphological stability of sharp 
interfaces \cite{Till63,MuSe64} has been used to examine the initial structures
formed during the directional solidification of binary alloys. A cellular growth during the
solidification is described by  {\sc Coriell} {\it et al.} \cite{CoFaSe85}. Growing sea ice at the 
ocean-atmosphere interface typically consists of an array of small ice platelets with randomly 
oriented horizontal c-axes which grow vertically downwards from the surface. 
{\sc Wettlaufer} \cite{We92} applied the morphological stability theory to the solidification of 
salt water and determined the length scales for ice platelet growth. Phase field methods have
improved the early interface stability approaches in terms of the time dependence of length scales 
and details of crystal shapes involved in pattern formation during directional solidification. 
However, the more simplistic sharp interface models have been shown to be capable to predict 
reasonable length scales within crystal growth \cite{Till91a,Till91b,Hur93}. 
The influence of natural convection \cite{WW97} has been shown to be included to properly 
predict the length scales at the ice-seawater interface \cite{Ma07}. 

When sea ice grows in turbulent surface water the newly formed platelets of ice are suspended 
in the water, forming the so-called frazil ice. This first form of new ice agglomerates by the 
action of wave motion and wind, and by further consolidation it forms larger aggregates such as 
grease ice and pancake ice. Finally a closed cover of granular ice is developed. This layer 
is characterized by the loss of prevalent orientation of the ice crystals due to their quick 
formation under turbulent water conditions. The underside of growing sea ice typically consists 
of a matrix of pure ice platelets which grow in the direction normal to the interface, forming 
the so-called skeletal layer. In this regime the platelets are generally parallel and the brine
circulates freely in vertical layers parallel to the platelets. Spacings between brine layers 
are typically $0.1-1$ mm, depending on the growth rate of the ice \cite{Weeks86}. Within the
skeletal layer a direct vertical connection to the seawater below exists. As the platelets 
lengthen and grow wider, small ice bridges form between adjacent platelets, forming the so-called
bridging layer. The formation of ice bridges, i.\ e.\ trapping inclusions of brine is the source 
of most of the liquid found in sea ice. Thus, above the skeletal layer, brine tends to be located
in vertically oriented sheetlike inclusions. 

{\sc Pringle} {\it et al.} \cite{Pring09} imaged sea ice single crystals with X-ray 
computed tomography, and found arrays of near-parallel intracrystalline brine layers whose
connectivity and complex morphology varied with temperature. 
The results of {\sc Pringle} {\it et al.} \cite{Pring09} clearly show near-parallel layering, 
but the pore space has been shown to be much more complicated than suggested by simple
models of parallel ice lamellae and parallel brine sheets (e.g. \cite{Weeks86}). While the
morphological instability at the ice-seawater interface is well understood, comparatively little 
is known, however, about the inclusions of brine due to the formation of ice bridges between the
growing ice platelets. Several interface stability approaches provided details of ice platelet
shapes, but they do not address pattern formation during the freezing of the liquid sandwiched
between ice platelet boundaries. In a step towards the development of a predictive model of sea 
ice microstrucrure, we develope a phase field model for the pattern formation during solidification 
of the two-dimensional interstitial liquid. We use the model to calculate the short-time frozen
microstructures and compare our findings with the vertical brine pore space obtained from 
X-ray computed tomography \cite{Pring09}.

In former work we have used the Turing mechanism to describe brine entrapment in sea ice
without salinity conservation \cite{KMG09}. In this paper we develop a theory that includes 
both macroscopic salt diffusion and microscopic order parameter dynamics. In such a so-called 
phase field model the two variables(salinity and order parameter) are coupled in two ways: 
salt is rejected from freezing water and accumulates in the water adjacent to the growing ice, 
and as the rejected salt causes the salinity of the seawater to increase, the free energy that
determines the order parameter dynamics is modified. In this paper we examine two phase field
equations which describe the time evolution of the non-conserved, Landau-Ginzburg order parameter
coupled to a spatially varying conserved field. Choosing the total mass of salt as the conserved
variable as done by {\sc Grandi} \cite{Grandi13} it follows a model in which the order parameter
is coupled to the local salinity. In fact, the author also developed a phase field approach for
the solidification and solute separation in water solutions. Both phase field approaches differ 
in the choice of the free energy functional. {\sc Grandi} \cite{Grandi13} started from a free
energy functional which is able to reproduce the equilibrium conditions for the coexistence of
ice and solute at critical concentration. We describe the free energy according to the classical
Landau-Ginzburg theory of a first-order phase transition, i.\ e.\ in terms of a power series
expansion of the free energy in the order parameter. The Landau coefficient functions we extract
from the thermodynamic properties of pure water in a way such that the equilibrium as well as the
metastable states of the ice/water system is well defined. One important difference is 
that {\sc Grandi} \cite{Grandi13} calculated the equilibrium phase profiles in one dimension,
whereas we consider the short-time frozen structures, which are non-equilibrium phases. Here we
develop a micrometer scale model where we neglect the spatial temperature gradient in ice or
water. Solving our phase field equations in one and two spatial dimensions we find that 
short-time frozen structures appear which are in a good agreement with the observed patterns.

The outline of the paper is as follows. In the next chapter we derive the model equations
from a variational principle. The linear stability analysis is presented in the
second chapter leading to the parameter range where structure can be formed. The numerical
solution and the comparison with the experimental data complete the fourth section. In the 
fifths section we summarize and conclude.

\section{Model development}

\subsection{Structure properties and order parameter of water and ice}

We restrict to  a hexagonal ice-I modification, the so-called
$I_h$, and search for a variable $u$ which minimizes the free energy
$\mathfrak{F}$. This variable $u$ should describe the
orderliness of water molecules as it was introduced
by {\sc Medvedev} {\it et al.} \cite{Me}. Ideal pure hexagonal ice possesses
an exact tetrahedral structure (see Fig. \ref{b0} below). Therefore the
measure of the state of order is the ''tetrahedricity' 
\begin{equation}
u \sim 1-M_T=1-\frac{1}{15 <l^2>}\sum_{i,j}(l_i - l_j)^2 ,
\end{equation}  
where $l_i$ are the lengths of the six edges of the tetrahedron
formed by the four nearest neighbors of the considered water molecule. For an
ideal tetrahedron one has $M_T=0$ and the random structure yields $M_T=1$. 

\begin{figure}[h]
\includegraphics[width=8cm]{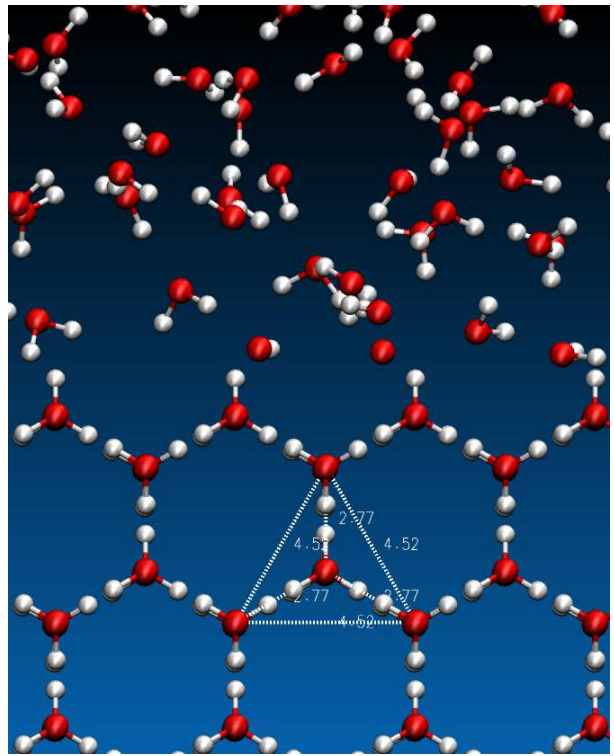}
\caption{\label{b0} 
Change from a regular hexagonal lattice structure to irregular bonds during 
the melting of ice: water at 300K (above) and hexagonal ice (below).
}
\end{figure}

In this way it is possible to discriminate between ice- and
water molecules via a two-state function. This corresponds to the idea of
{\sc Wilhelm  Conrad R\"ontgen}, who  distinguished between molecules of the
first kind, which he called ice molecules and molecules of the second
kind \cite{Roe} representing the liquid aggregate state. 

The equations of the model we derive from non-conserved, Landau-Ginzburg
order parameter kinetics together with salinity conservation assuming the 
total free energy in d-dimensions
\begin{equation}
\mathfrak{F} = \int\left\{\frac{1}{2}K\left [\nabla u(x,t) \right]^2 +
  f[u(x,t),v(x,t)]\right \}d^dx 
\label{free_energy}
\end{equation}
with the energy density for the order parameter $u$ and the salinity $v$
\begin{equation}
f(u,v) = \frac{a_1'}{2}u^2 -  \frac{a_2}{3}u^3 +
  \frac{a_3}{4}u^4  + \frac{h}{2}u^2v + \frac{b_1}{2}v^2-b_0v .
\label{Functional}
\end{equation}
The parameter $a_1'$ is the temperature-dependent freezing parameter determining the phase
transition. 
The temperature dependence of the cubic and quartic terms of the Landau-Ginzburg-functional 
(\ref{free_energy}) is supposed to be so weak near the phase transition that 
they can be well approximated as positive constants, namely, $a_2$ and $a_3$. The structure
parameter $a_3$ is responsible for the stability of the ordered phase, $K$ is the gradient energy
coefficient of order parameter fluctuations, and $b_1$ determines the salinity fluctuations. 
The coefficient $a_2$ is connected with an uneven exponent and is therefore responsible for 
possible phase transitions of first kind. The coefficient $h$ describes the interaction between 
the order parameter $u$ and the salinity $v$.
We will see soon that these parameters can be scaled to only three relevant parameters where 
the phase diagram is determined only by two of them, the dimensionless structure and 
freezing parameter.

The variation of the Landau-Ginzburg free energy (\ref{free_energy}) yields
the gradient dynamics for the order parameter $u$ 
 \begin{equation}
\frac{\partial u}{\partial t}= 
-\Gamma\frac{\delta  \mathfrak{F}}{\delta u}= \Gamma\left(
-a_1'u + a_2 u^2 - a_3u^3  - huv + K\Delta u\right)\, ,
\label{lg_1}
\end{equation}
where $\Gamma$ is a relaxation coefficient for $u$ which determines the time scale of the 
ordering process. Further we want to assume that the total mass of salt is conserved. Therefore
we demand to have a balance equation of the form $\partial v/\partial t=-\nabla \vec j$
where the current is assumed to be proportional to a generalized 
force $\vec j=M\vec f$ which should be given by a potential as $\vec f=-\nabla P$. 
This potential in turn is given by the variation of the free energy 
density $P=\delta {\mathfrak{F}}/\delta v$. This procedure is nothing but the the second law
of Fick. With the assumption that salt is a passive tracer which does not undergo a phase
transition, we obtain an equation of Cahn-Hilliard-type without the fourth derivation for 
the evolution of the salinity $v$       
\begin{equation}
\frac{\partial v}{\partial t}=
\nabla\left(M\nabla\frac{\delta \mathfrak{F}}{\delta v}\right)= 
M\left(\frac{h}{2}\Delta u^2 + b_1\Delta v\right) . 
\label{lg_2}
\end{equation}
The coefficient $M$ is the mobility, which is considered as constant. This linearisation
of the equation is valid for the early phase of brine entrapment where the brine pore 
space is connected. 
 
Defining the reduced time and spatial coordinates 
\be
\tau=\frac{\Gamma b_1 a_2^2}{h^2}t,\qquad 
\xi=\sqrt{\frac{\Gamma}{M}}\frac{a_2}{h}x
\ee
as well as dimensionless order parameters of ice/water structure and salinity
\be
\psi=\frac{h^2}{b_1 a_2}u,\qquad
\rho=\frac{h^3}{b_1 a_2^2}v
\ee
the model equations (\ref{lg_1}) and (\ref{lg_2})
can be written in reduced form
\begin{eqnarray}
\frac{\partial \psi}{\partial \tau} &=& -\alpha_1'\psi + \psi^2 -
\alpha_3\psi^3 - \psi\rho + D\frac{\partial^2 \psi }{\partial \xi^2} 
\label{f1} \\
\frac{\partial \rho}{\partial \tau} &=& 
\frac{1}{2}\frac{\partial^2 \psi^2}{\partial \xi^2} + 
\frac{\partial^2 \rho}{\partial \xi^2}. 
\label{f2}
\end{eqnarray}
The benefit of this form is that the dynamics of the dimensionless
order parameter $\psi$ and the dimensionless salinity $\rho$ depends
only on three parameters, the freezing parameter
\begin{equation}
\alpha_1'=\frac{a_1'}{a_2^2}\frac{h^2}{b_1}\, ,
\label{freezp}
\end{equation}
the structure parameter
\begin{equation}
\alpha_3=\frac{a_3 b_1}{h^2}\, ,
\label{structurep}
\end{equation}
and the diffusivity $D=\frac{D_{\rm ice}}{D_{\rm salt}}$  with $\alpha_1,'\,\alpha_3,\,D > 0$.
The diffusivity $D_{\rm ice}=\Gamma K$ describes the propagation of the order parameter, 
and $D_{\rm salt}=M b_1$ the diffusion of salt. The parameters $\alpha_1'$ and $\alpha_3$ 
define the regions for the ordered and non-ordered phase. Their physical meaning will be 
more transparent when the results are discussed in terms of thermodynamic properties of water.

\section{Linear stability analysis}

\subsection{Uniform stationary solution}

The following linear stability analysis is related to one spatial
dimension. The extension to higher dimensions is straight forward.

In terms of the reduced quantities, the free energy density (\ref{Functional})
for a uniform order parameter and salinity  
\begin{equation}
\psi = \psi_0 = const \, , \quad \rho = \rho_0 = const
\end{equation}
takes the form
\begin{equation}
f(\Psi_0,\rho_0) = \alpha_0+\frac{\alpha_1}{2}\psi_0^2-\frac{1}{3}\psi_0^3+\frac{\alpha_3}{4}\psi_0^4
\end{equation}

where $\alpha_1$ and $\alpha_0$  depend on the 
salinity $\rho_0$ as
\begin{eqnarray}                
\alpha_0 & = & \frac{1}{2}\rho_0^2 - \gamma \rho_0
\nonumber \\
\alpha_1 & = & \alpha_1' + \rho_0 \, 
\label{fz}.
\end{eqnarray}

Important for the further description of seaice formation is the compound
parameter $\alpha_1$. The sum (\ref{fz}) includes both the
temperature-dependent parameter $\alpha_1'$ and the salinity
$\rho_0$. If the parameter $\alpha_1'$ corresponds to a certain freezing point 
temperature then $\alpha_1'+\rho_0$ corresponds to a lower transition temperature. 
This effect is known as freezing-point depression. 

Note that $\alpha_0$ is the value of $f$ in the disordered phase, which is
characterized by $\psi_0 = 0$. Since $\alpha_0$ is a constant background term,
which is merely dependent on the total salinity, it can be taken to be zero
for the subsequent discussion, i.e.
\begin{equation}
f(\Psi_0,\rho_0) = \psi_0^2(\frac{\alpha_1}{2} - \frac{1}{3}\psi_0 + \frac{\alpha_3}{4}\psi_0^2).
\label{fzdiff}
\end{equation}

\begin{figure}[]
 \includegraphics[width=8cm]{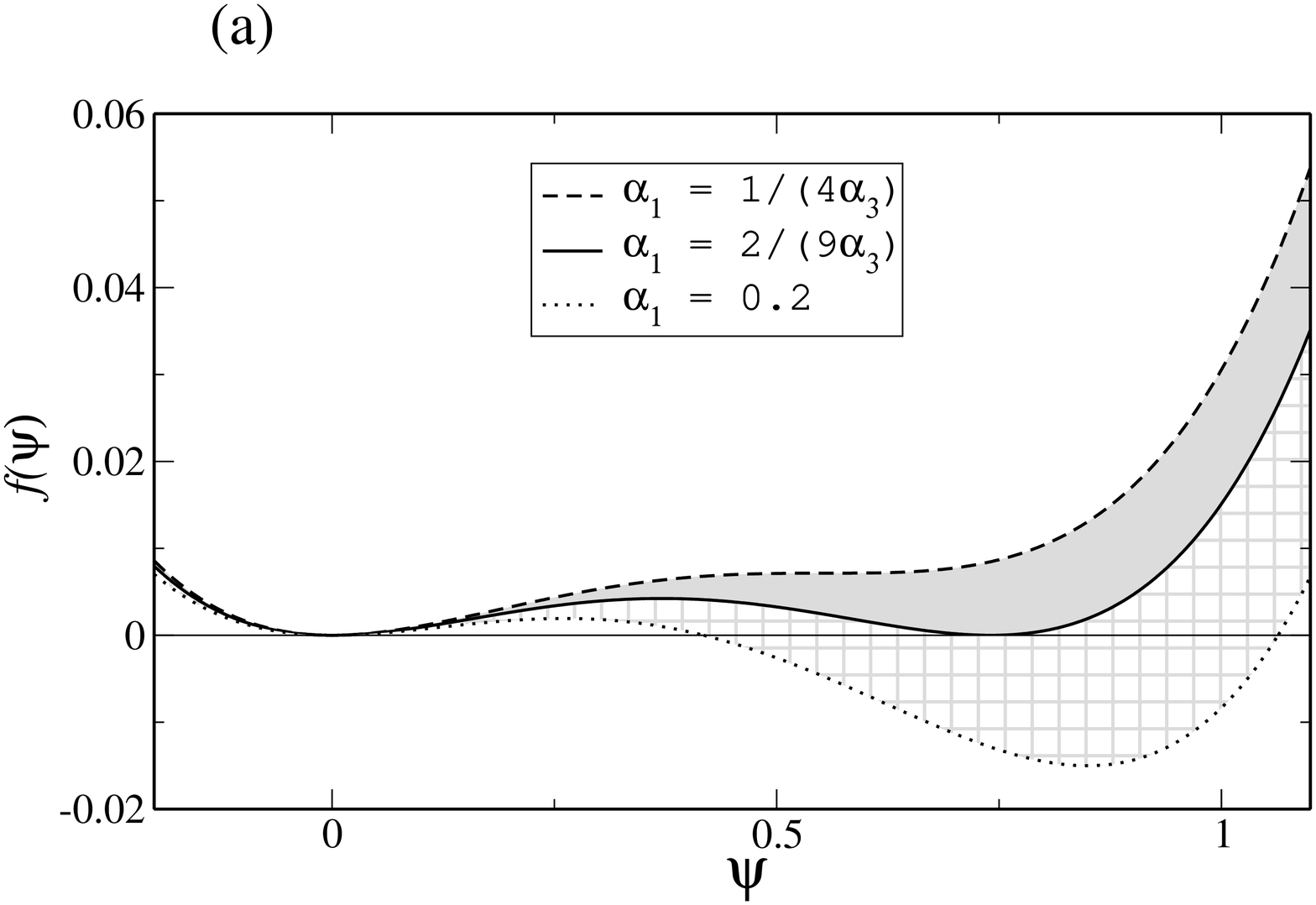}
 \includegraphics[width=8cm]{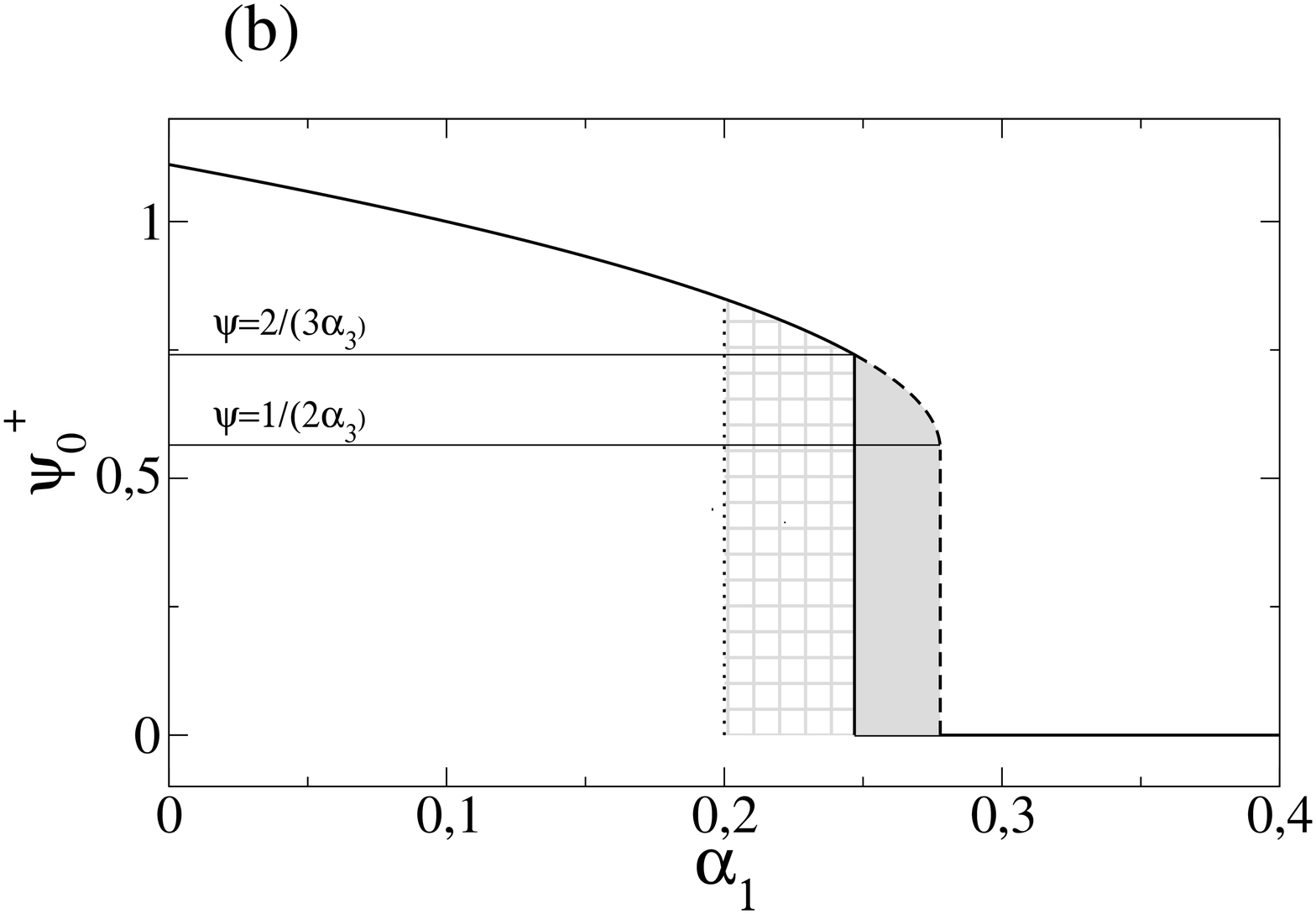}
\caption{Condition for a first-order phase transition. 
(a) The free energy density $f$ versus the uniform dimensionless 
order parameter (tetrahedricity) for various freezing parameter $\alpha_1$ and
the structure parameter $\alpha_3=0.9$. 
(b) Dependence of absolute minimum of the free energy density (\ref{psi0p}) on $\alpha_1$
(solid line).  The dashed line corresponds to the position of the second
minimum and the dotted line indicates if one jumps from the first minimum to
the absolute one in figure (a).
In the checked-pattern region a first-order phase transition can occur and the ordered 
phase $\psi_0^{+}$ is stable. In the gray region, the ordered phase $\psi_0^{+}$ 
is metastable whereas the non-ordered phase $\psi_0 $ is stable.}
\label{FreeEnergy}
\end{figure}

The function $f$ versus the dimensionless order parameter for 
various $\alpha_1$ is shown in Fig. \ref{FreeEnergy}(a). The locations
of the extrema of $f$ are given by the stationary  condition
\begin{equation}
\frac{\partial f}{\partial \psi_0} = 0 = \psi_0(\alpha_1 - \psi_0 + \alpha_3\psi_0^2)
\end{equation}
whose solutions have a minimum (assuming $\alpha_1>0$)
\begin{equation}
\psi_0^0 = 0
\end{equation}
 and a minimum/maximum for
\begin{equation}
\psi_0^{\pm} = \frac{1}{2\alpha_3}\left (1 \pm
\sqrt{1-4\alpha_1\alpha_3}\right ).
\label{psi0p}
\end{equation}
At sufficient high $\alpha_1>1/4\alpha_3$, the minimum at $\psi_0^0 = 0$ is the only 
allowed physical solution, which is the disordered state.  As soon as 
\be
\alpha_1 \le {1\over 4\alpha_3}
\label{inst}
\ee 
a second relative minimum appears at $\psi_0^{+}$.  Which of these minima establishes the 
state of lowest free energy is dependent on the critical $\alpha_{1,c}$ which is found from 
the coexistence curve where these two local minima are equal 
and $f(\Psi_0^+)=f(\Psi_0^0)= 0$ which yields
\be
\alpha_{1,c}={2\over 9\alpha_3}.
\label{16}
\ee
The coexistence curve is plotted as solid line in Fig. \ref{FreeEnergy}(a) and 
its second minima becomes $\psi_0^+(\alpha_{1,c})=\frac{2}{3\alpha_3}$.
Above the critical parameter $\alpha_{1,c} < \alpha_1 < \frac{1}{4\alpha_3}$ the ordered 
phase $\psi_0^{+} > 0$ is metastable whereas the non-ordered phase ($\psi_0 =
0$) is stable. This is illustrated as gray area in
Fig. \ref{FreeEnergy}(a). As soon as $\alpha_1$ decreases approaching
$\alpha_1 = \alpha_{1,c}$ the second  minimum at $\psi_0^{+}>0$ becomes deeper
than the minimum at $\psi_0 = 0$ and the ordered phase $\psi_0^+$ becomes stable. 
This is illustrated as checked pattern area. 
Therefore the absolute minimum is dependent on a discontinuous change 
of the order parameter from $\psi_0 = 0$ to $\psi_0^{+} > 0$ as plotted
in Fig. \ref{FreeEnergy}(b).

\subsection{Conditions for structure formation}

The fixed points for the kinetics (\ref{f1}) and (\ref{f2}) 
are given by the local extrema of the Landau function $f$ as discussed in the foregoing 
paragraph. In order to derive the conditions for structure formation, we perform
a linear stability analysis for the two local minima around 
the disordered phase $\psi_0^0$ and the ordered phase $\psi_0^{+}$ by linearizing the equation 
system (\ref{f1}) and (\ref{f2}) according 
to $\psi=\psi_0 + \bar \psi$ and $\rho=\rho_0 + \bar \rho$. 
For the time evolution of the linear perturbations $\bar \psi$ 
and $\bar \rho$ one gets the equation system
\begin{eqnarray}
 \begin{pmatrix}
   {\partial \bar\psi \over \partial{\tau}}
    \\[1ex]
   {\partial \bar\rho \over \partial{\tau}}  
\end{pmatrix} &=&   
\begin{pmatrix}
{\aleph} +D{\partial^2 \over \partial \xi^2}& -\psi_0 \\[1ex]
 \psi_0 {\partial^2 \over \partial \xi^2} &   {\partial^2 \over \partial \xi^2} 
\end{pmatrix} 
\begin{pmatrix}
 \bar\psi \\[1ex]
 \bar\rho  
 \end{pmatrix}  \, .  
\label{linearis}
\end{eqnarray}
Here ${\aleph}=-\alpha_1+2\Psi_0-3 \alpha_3 \Psi_0^2$ which takes the value
${\aleph}=-\alpha_1$ for the fixed point $\Psi_0^0=0$ and ${\aleph}=\psi_0-2\alpha_3\psi^2_0$
for $\Psi_0^\pm$.

As commonly used, the Fourier ansatz 
$ \bar \rho = \rho_0  \exp[\lambda(\kappa) \tau +i\kappa\xi]$
and analogously for $\psi$ results into the characteristic equation for the growth 
rate $\lambda(\kappa)$
\begin{equation}
 \lambda(\kappa)^2 +\left[(D+1)\kappa^2-{\aleph}\right]\lambda(\kappa)+q(\kappa^2)=0
\label{lam}
\end{equation}
 with
\begin{equation}
 q(\kappa^2) = \kappa^2\left[D\kappa^2 -{\aleph}- \psi^2_0\right].
\label{qkappa}
\end{equation}
The two possible growth rates read 
\begin{eqnarray}
\lambda_{1,2} = &-& \frac{1}{2}\left [(D+1)\kappa^2 -{\aleph}  \pm
  \sqrt{\Delta}\right ]
\label{rootslam}
\end{eqnarray}
with
\begin{eqnarray}
\Delta & = &  
 [(D+1)\kappa^2 -{\aleph}]^2 - 4 q(\kappa^2) \nonumber \\ 
          & = & 
 [(D-1)\kappa^2 -{\aleph}]^2 + 4 \kappa^2\psi^2_0 > 0.
 \label{diskrim}
\end{eqnarray}

Time-oscillating structures can appear only if ${\rm Im}\ \lambda(\kappa)\neq 0$, 
i.e. $\Delta<0$, which can never be  fulfilled
since (\ref{diskrim}) can be expressed as a sum of two real squared numbers.
In other words we do not have time-growing oscillating structures in our model. 

The condition for structure formation is the instability of the fixed point, 
where some spatial fluctuations may be amplified and form macroscopic structures. 
An unstable fixed point which is associated with 
positive eigenvalues $\lambda(\kappa)>0$ allows any fluctuation 
with a wave-vector $\kappa$ to grow exponentially in time.  
Therefore, we search for eigenvalues which possess a positive real part for a 
positive wave number, i.e. ${\rm Re}\ \lambda(\kappa)>0$ and $\kappa^2>0$. 

\begin{figure}[]
\includegraphics[width=9cm]{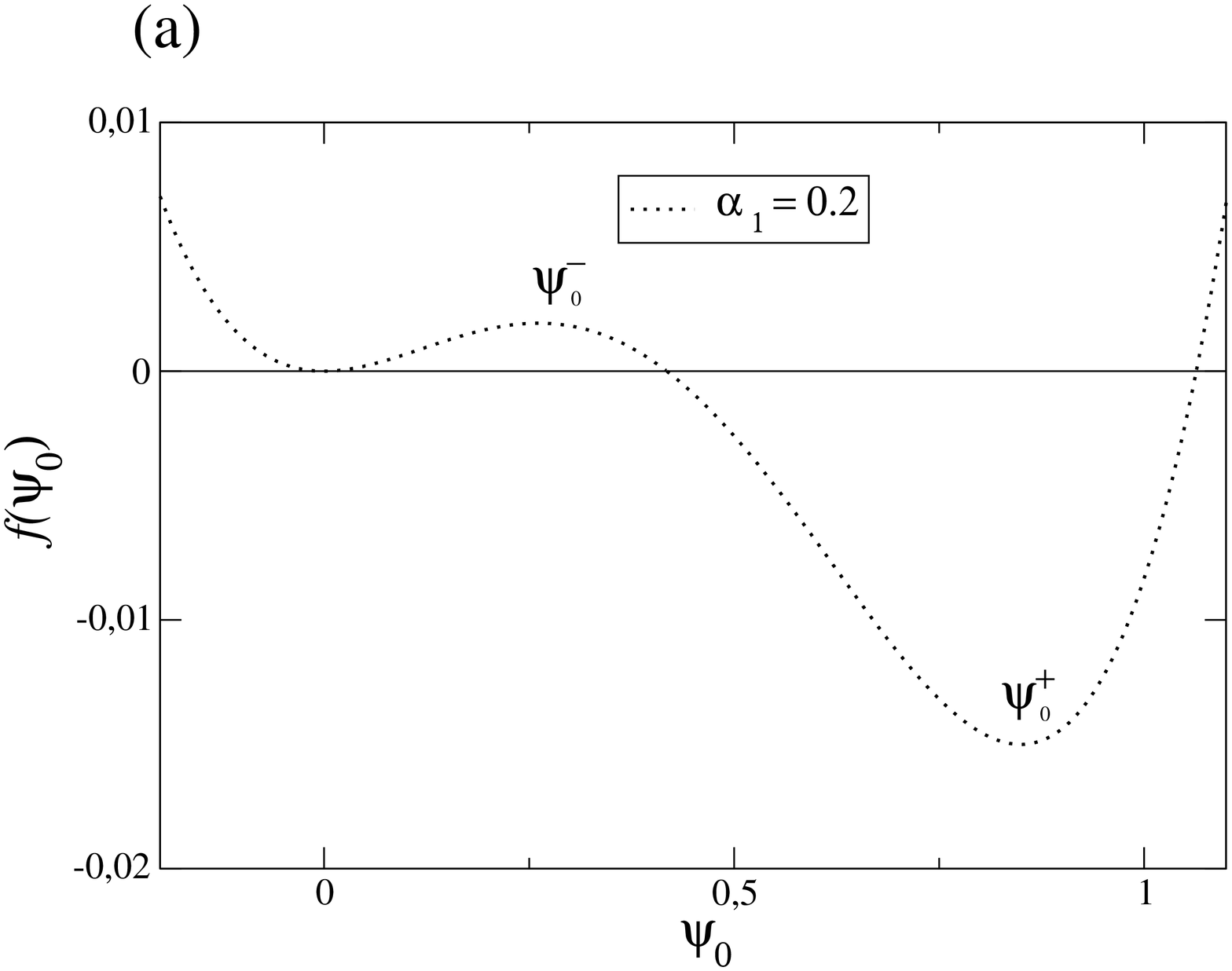}
\includegraphics[width=9cm]{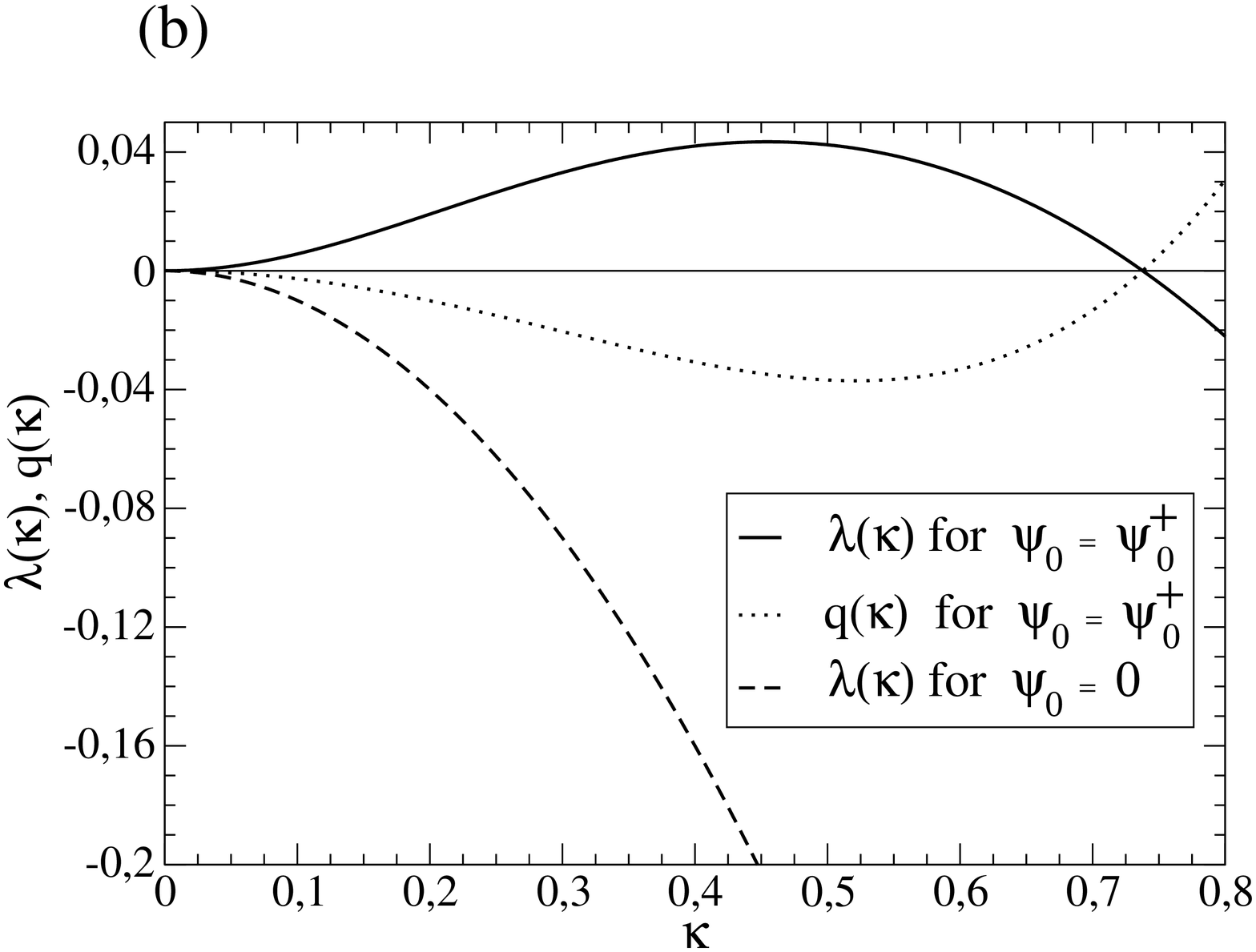}
  \caption{Conditions for structure formation. 
(a) The free energy density $f$ versus the uniform dimensionless 
order parameter for $\alpha_1  < \alpha_{1,c}$.
(b) Eigenvalues $\lambda(\kappa)$ versus the dimensionless wave number $\kappa$ 
for $\alpha_3 = 0.9$, $\alpha_1 = 0.2$, and $D = 0.5$ along 
with the function $q(\kappa^2)$ of equation (\ref{qkappa}).}
\label{EigenVal}
\end{figure}

The condition for structure formation is not satisfied for each fixed point. 
For the fixed point representing the disordered phase, $\psi_0 = 0$ and $\rho_0=const$, 
one sees from (\ref{rootslam}) that there are no positive real 
eigenvalues
\be
\lambda_{1,2}=\frac 1 2 \left [ -(D+1) \kappa^2-\alpha_1\pm
  |(D-1)\kappa^2+\alpha_1|\right ]<0,
\ee
see also Fig. \ref{EigenVal}. 
Consequently, there is no structure formation in this state which was expected
for the disordered phase, of course.

Now lets concentrate on the fixed point for the ordered phase $\Psi_0^+$ of (\ref{psi0p}). 
There are two distinct roots (\ref{rootslam}), both of which
are real numbers. For the condition of time-growing fluctuations we search now for the 
condition $\lambda(\kappa)>0$. The coefficient of the linear term of equation (\ref{lam}) is positive 
since $-{\aleph}=2\alpha_3\psi^2_0 - \psi_0 > 0$ for $\psi_0 = \psi_0^{+} > 1/(2\alpha_3)$ 
and $\kappa^2$ as well as $D > 0$. Therefore, the first term in (\ref{rootslam}) is negative and
we can only have a positive $\lambda(\kappa)>0$ if the square of this term is less than the 
discriminant $\Delta$. This leads to the condition $q(\kappa^2) < 0$. The latter is illustrated for the
parameters $\alpha_3 = 0.9$, $\alpha_1 = 0.2$, and $D = 0.5$ in Fig. \ref{EigenVal}, which shows $\lambda(\kappa)$ 
versus the dimensionless wave number $\kappa$ along with $q(\kappa^2)$. 

\begin{figure}[ht]
 \includegraphics[width=9cm]{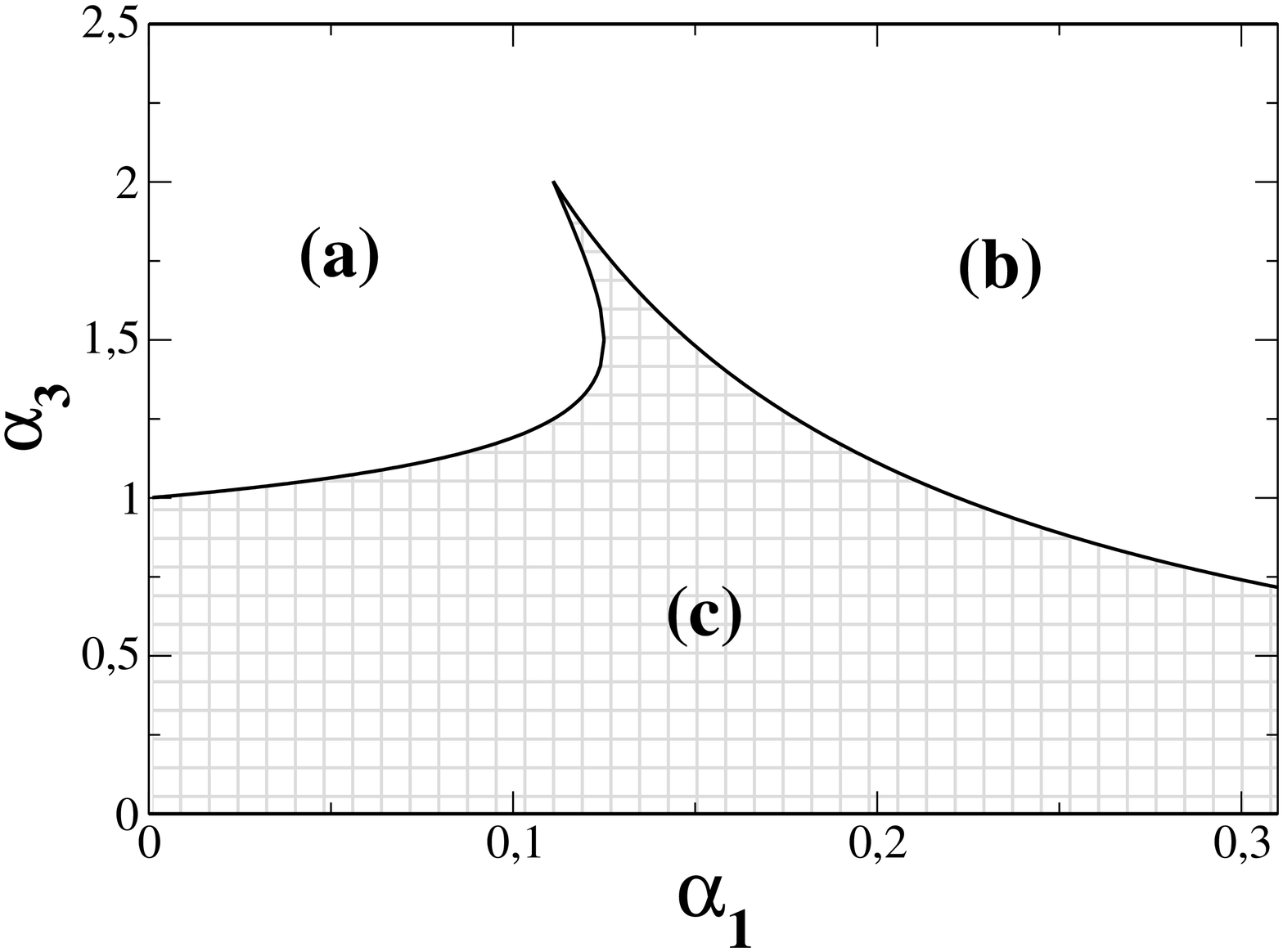}
\parbox[]{9cm}{
\parbox[]{4cm}{
(a)\\ 
\includegraphics[width=3.5cm]{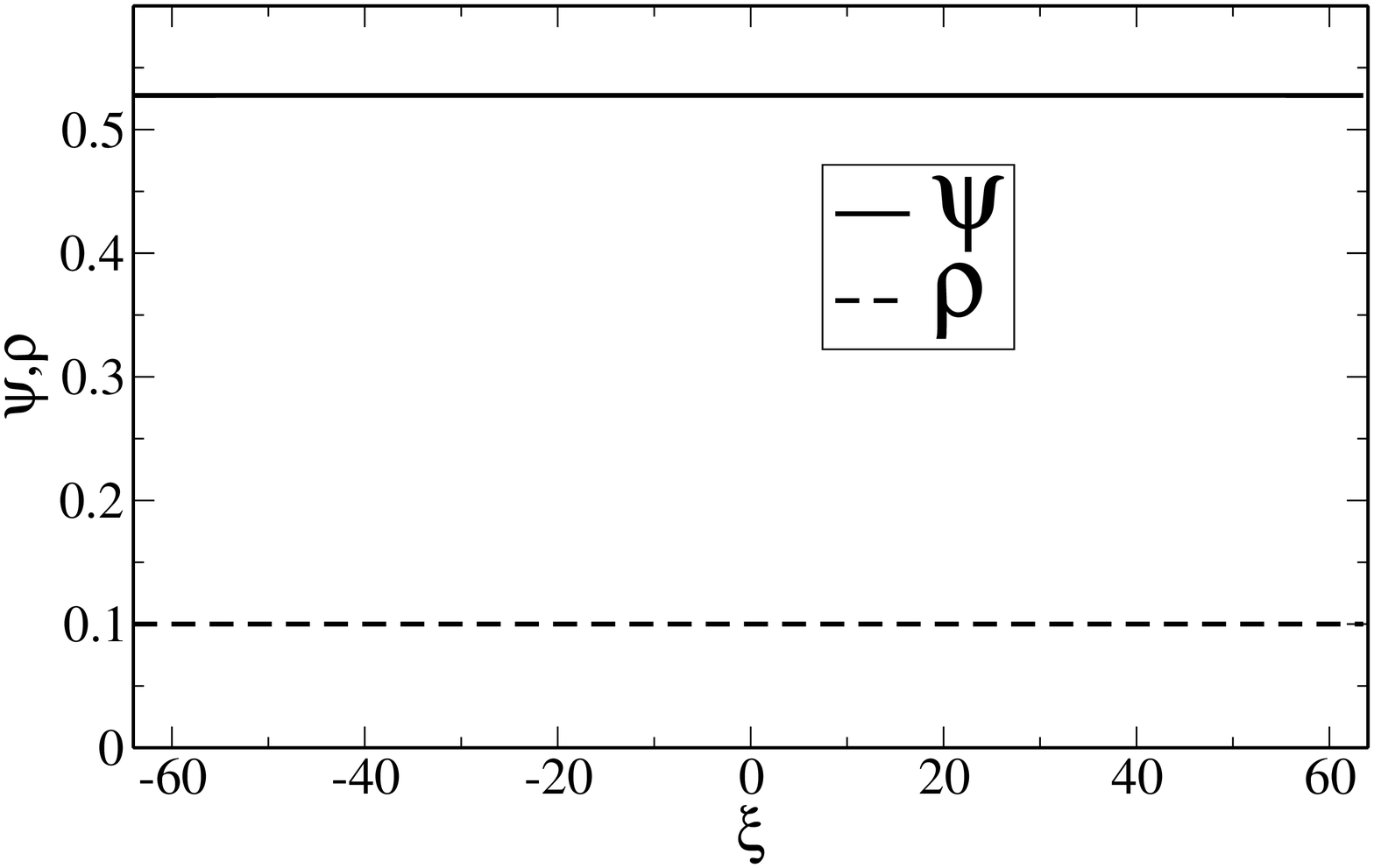}
}
\parbox[]{4cm}{
(b)\\ 
\includegraphics[width=3.5cm]{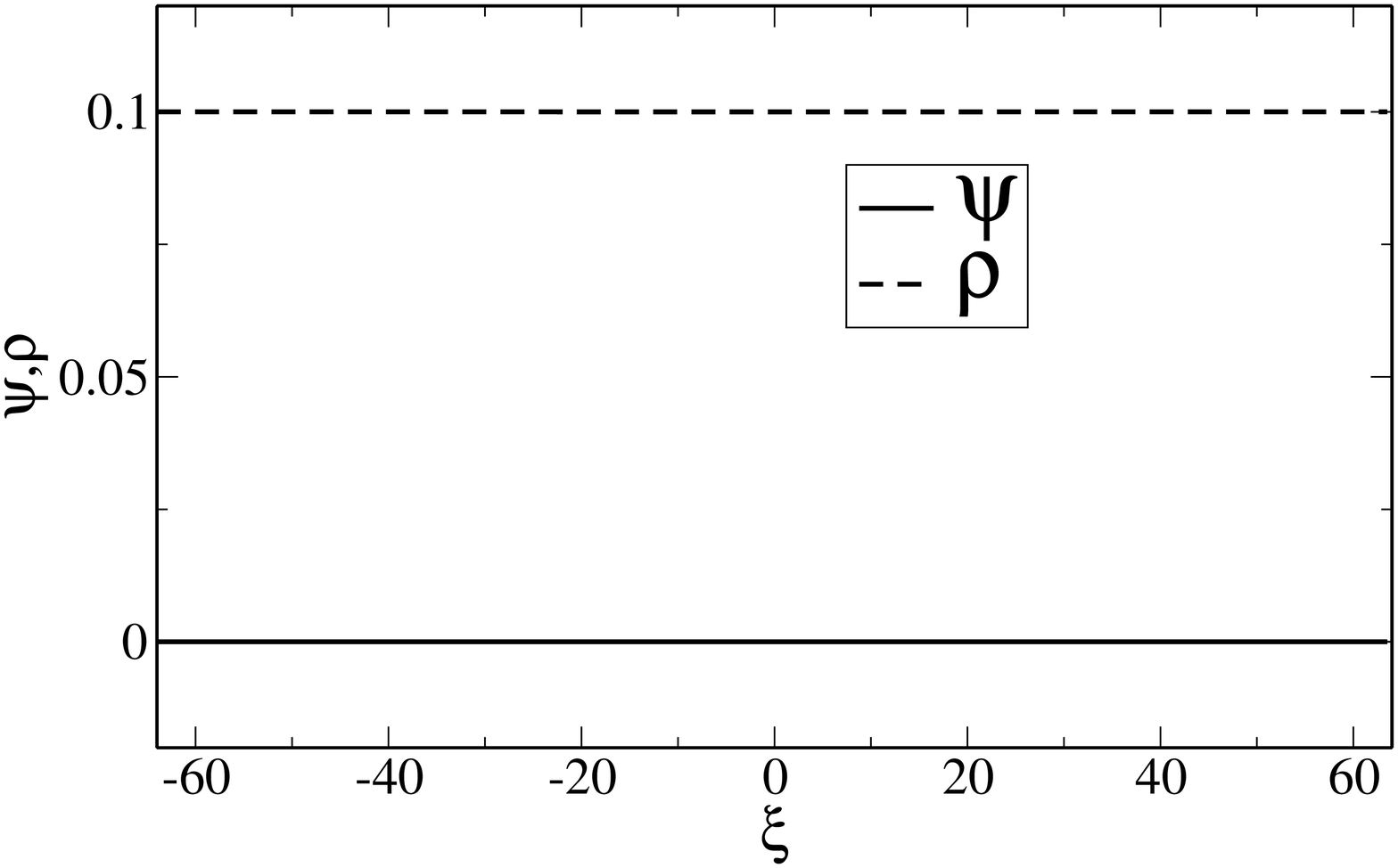}
}
}
\parbox[]{9cm}{
(c)\\ 
\includegraphics[width=4cm]{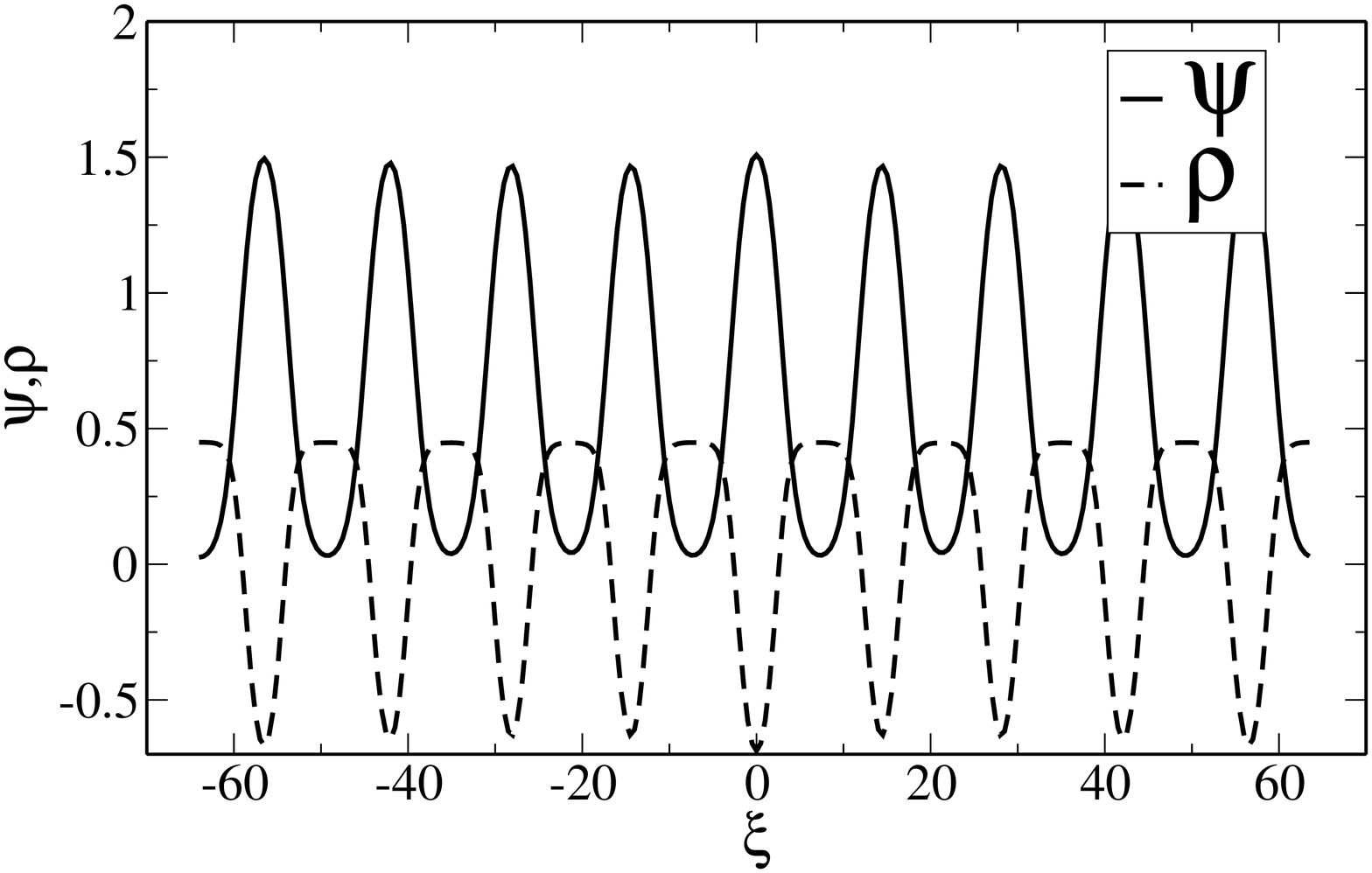}
}
\caption{The instability region of the fixed 
 point $\psi_0^{+}$ of (\ref{psi0p}) and $\rho_0 = const$ as phase
diagram together with selected structures: (a) $\alpha_1'=0.01$,
 $\rho(\tau=0)=0.1$, $\alpha_3=1.5$, (b) $\alpha_1'=0.2$, $\rho(\tau=0)=0.1$, $\alpha_3=0.9$, (c)
 $\alpha_1'=0.1$, $\rho(\tau=0)=0.1$, $\alpha_3=0.9$. In the checked region spatial structures can occur.}
\label{InstabReg}
\end{figure}

We can only have 
positive $\lambda(\kappa)$ if the values of $\kappa$ are restricted to the region between the zeros 
of $\lambda(\kappa)$, which coincide with the roots of the equation
\begin{equation}
 q(\kappa^2) = [-\psi_0 + (2\alpha_3 - 1)\psi^2_0 + D\kappa^2]\kappa^2 = 0 \, .
\end{equation}
 Hence, the $\kappa$ region is restricted to the interval
\begin{equation}
 \kappa^2\in \left(0,\,{\psi_0 \over D} \left[1 - (2\alpha_3 - 1)\psi_0\right]\right)
 \label{cond1a}
\end{equation}
 with $\psi_0=\psi_0^{+}$. The upper boundary of the interval must be positive, 
and we get only a meaningful condition from (\ref{cond1a}) if 
$\psi_0\left[1 - (2\alpha_3 - 1)\psi_0\right]/D > 0$.  Discussing separately the cases $\alpha_3\gtrless 1/2$ one sees that (\ref{cond1a}) gives no additional restriction 
on $\alpha_1$ and $D$ for $\alpha_3 \leq 1/2$. For the case $\alpha_3 > 1/2$ we get
the restriction
\begin{equation}
 (2\alpha_3 -1)\sqrt{1-4\alpha_1\alpha_3} < 1 \, , 
 \quad \alpha_1 > 0 \, , \quad \alpha_3 > 0 \, .
 \label{cond1b}
\end{equation}

Moreover, the latter one has to be in agreement with the condition for the occurrence of a 
first-order phase transition (\ref{16}) 
\begin{equation}
 0 < \alpha_3 < \frac{2}{9\alpha_1}\, .
 \label{cond1c}
\end{equation}

 Solving the set of inequalities given by (\ref{cond1b}) and (\ref{cond1c}),
 we obtain the range for structure formation
\be
 2 > \alpha_3 >1:&& {1\over 4 \alpha_3}\left (1-{1\over (2 \alpha_3-1)^2}\right
 ) <\alpha_1<\frac{2}{9\alpha_3}\nonumber\\
 1 > \alpha_3 >0:&& 0<\alpha_1<\frac{2}{9\alpha_3}
 \label{reg}
\ee
In Fig. \ref{InstabReg} we present this in form of a phase diagram for the
freezing and structure parameters. One can see that the instability region
starts at the maximal point $\alpha_1 = 1/9$ at $\alpha_3 = 2$
which means that we have only a structure formation for sufficiently 
small structure parameters $\alpha_3 < 2$ and proper bound on the freezing
parameter $\alpha_1$, see (\ref{reg}). 

The description of the instability region does not involve a restriction on 
the diffusivities of salt and water.
This is different from the model of {\sc Kutschan} {\it et al.} \cite{KMG09}, which
describes structure formation in seaice in terms of Turing structures.
The latter can only exist if the difference between the diffusion coefficients
of salt and water is sufficiently large. In the present study, the 
diffusivity $D$ enters the range for the possible wave 
numbers of unstable modes only.  According to (\ref{cond1a}) it is given explicitly by 
\ba
\kappa^2\in 
\left [0, 
{\alpha_1\over D}\left (2-{1\over \alpha_3}\right )+
  {(1-\alpha_3)\left(1+\sqrt{1-4\alpha_1\alpha_3}\right)\over
  2 D \alpha_3^2} 
\right ]\, .
\label{condk}
\end{align}
Fig. \ref{InstabReg} shows the influence of the structure parameter 
$\alpha_1$ on the formation of a solidification microstructure. 
A small $\alpha_1$ means low temperatures and/or low
salinities and consequently a freezing process. Therfore we find in the region
(a) the uniform ice phase and a inclusion of salt as a function of the
structure parameter $\alpha_3$. In contrast in the region (b) there are higher
temperatures and/or higher salinities. This induces a melting with
a uniform liquid water phase and dissolved salt. Finally the spatial
structures can appear in region (c).

\subsection{Mechanism of phase separation}

The formation of a spatial structure is driven by the instability of
the uniform stationary solution $\psi_0 = \psi_0^{+}$ and $\rho_0=const$,
as discussed in the previous chapter. The instability becomes evident in 
terms of a positive eigenvalue $\lambda(\kappa)$ for a range of 
wave numbers $\kappa$ as seen in Fig. \ref{EigenVal}. Hence, any linear 
perturbation with a wave number $\kappa$ inside the region between 
the zeros of $\lambda(\kappa)$ grows exponentially with amplification
factor $\lambda(\kappa)$. The function $\lambda(\kappa)$ has a maximum at 
a critical wave number $\kappa_c$, which defines the fastest-growing wave-vector

\begin{eqnarray}
\kappa^2_c & = & {\psi_0\over (D-1)^2}\biggl [ {(D-1) \left(1-2 \alpha_3 \psi_0\right)
                - 2 \psi_0}\nonumber   \\
           &+&  {(D+1)\psi_0^{1/2} \sqrt{(D-1) (2 \alpha_3
               \psi_0-1)+\psi_0}\over \sqrt{D}}\biggr ]\, 
\label{kcrit}
\end{eqnarray}
with $\psi_0 = \psi_0^{+}$ of (\ref{psi0p}). The expression for $\kappa_c$ 
can be approximated by the location of the minimum of the function 
$q(\kappa^2)$ in equation (\ref{qkappa}) 
\begin{equation}
 \kappa^2_{c} \approx \frac{\left(1-2 \alpha_3\right)\psi_0^2 + \psi_0}{2 D}\, .
 \label{kcappr}
\end{equation}
  
The critical wave number sets the length scale on which phase 
separation occurs. The size of the initial structure can be estimated
by $2\pi/\kappa_c$. The size of the structure depends on the 
parameters $\alpha_1$, $\alpha_3$, and $D$ as seen in (\ref{kcrit}) or (\ref{kcappr}). With the parameters chosen in
Fig. \ref{EigenVal}, we obtain a pattern size of $13.81$. Let us note already
here that we have chosen these parameters according to the physical properties of water as will be considered in chapter IV.

\begin{figure}[]
\includegraphics[width=8.5cm]{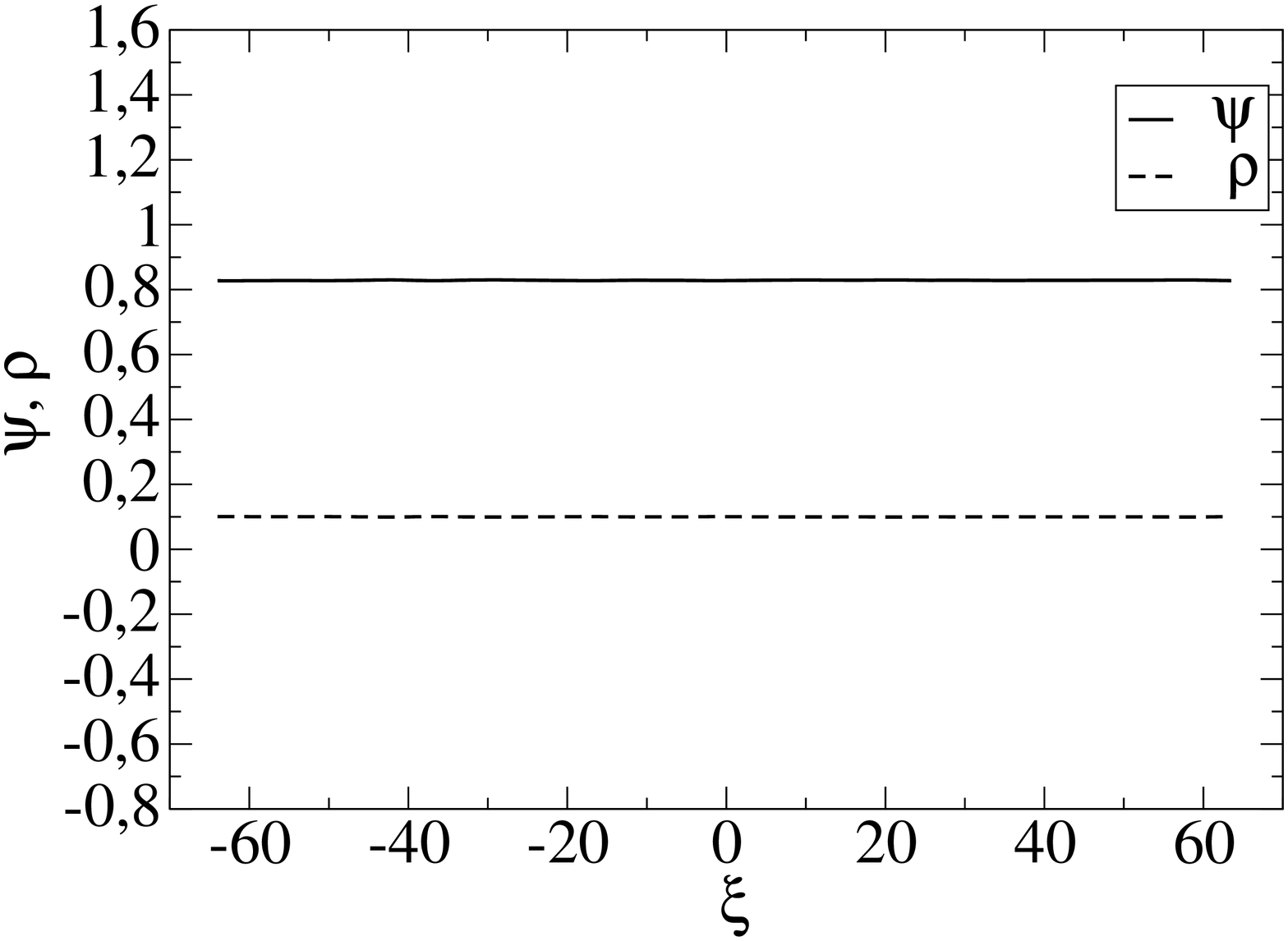}
\includegraphics[width=8.5cm]{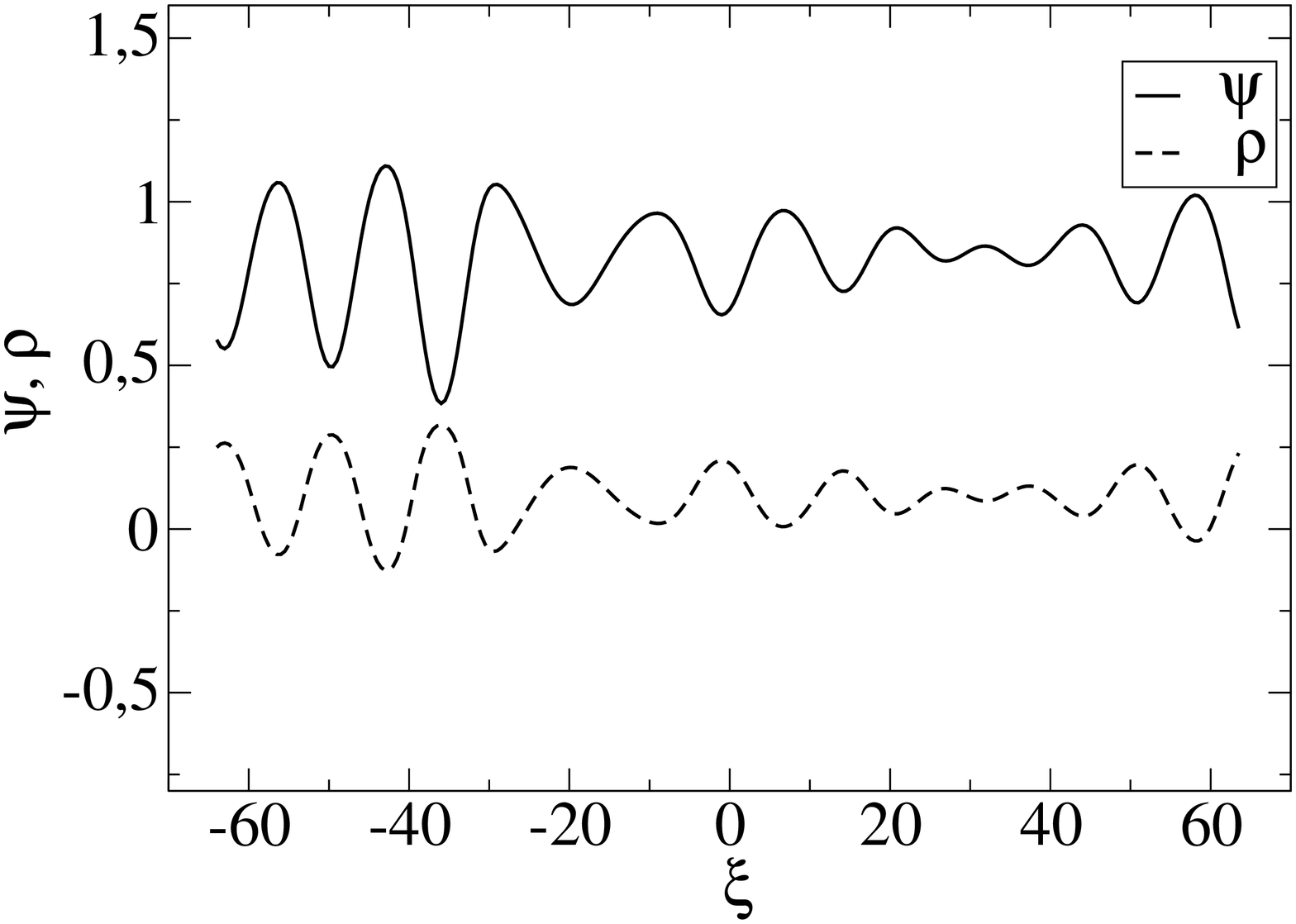}
\includegraphics[width=8.5cm]{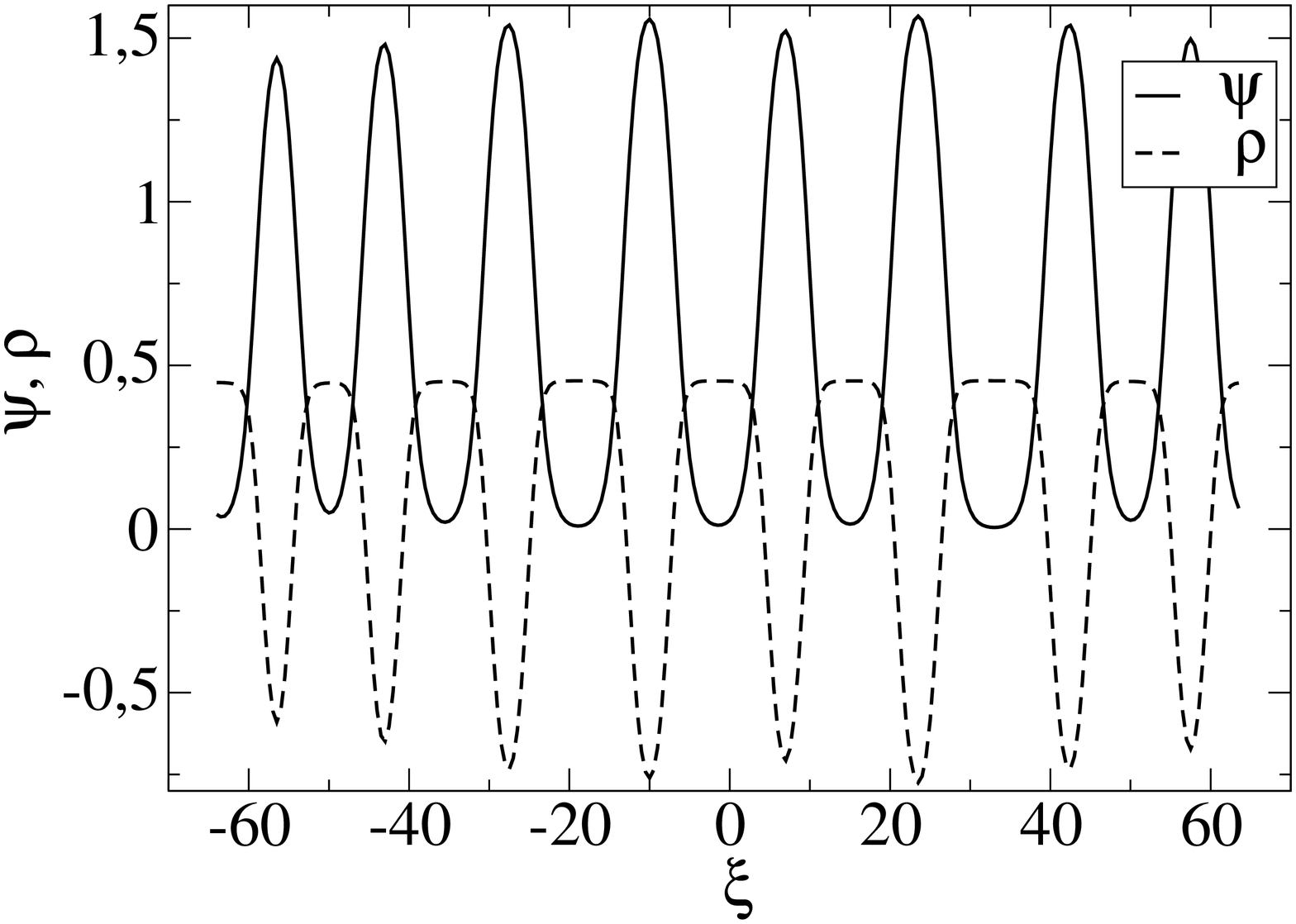}
  \caption{Time evolution of the order parameter
  $\psi$ and salinity $\rho$ versus spatial coordinates for    
  $\tau=10, 150, 500$ (from above to below) 
  for $\alpha_3 = 0.9$, $\alpha_1' = 0.1$, and $D = 0.5$ 
  with the initial conditions $\psi(\tau=0)=0.9$ and a randomly distributed 
  $\rho(\tau=0)=0.1 \pm 0.001$.}
\label{Num1D}
\end{figure}

\subsection{Numerical solution}

In order to proof the relevance of the linear stability analysis of the chapters above and 
to calculate the time evolution of phase separation, we now solve the equation 
system (\ref{f1}) and (\ref{f2}) numerically in one and two space 
dimensions. For this purpose we use the so-called exponential time
differencing scheme of second order (ETD2) \cite{CoMa} 
with the help of which a stiff
differential equation of the type $\dot y=r y+z(y,t)$ with a linear term $r
y$ and a nonlinear part $z(y,t)$ can be solved.  
The linear equation is solved formally and the integral over the nonlinear
part is approximated by a proper finite differencing scheme. 

\begin{figure}[h]
  \includegraphics[width=8cm,angle=0]{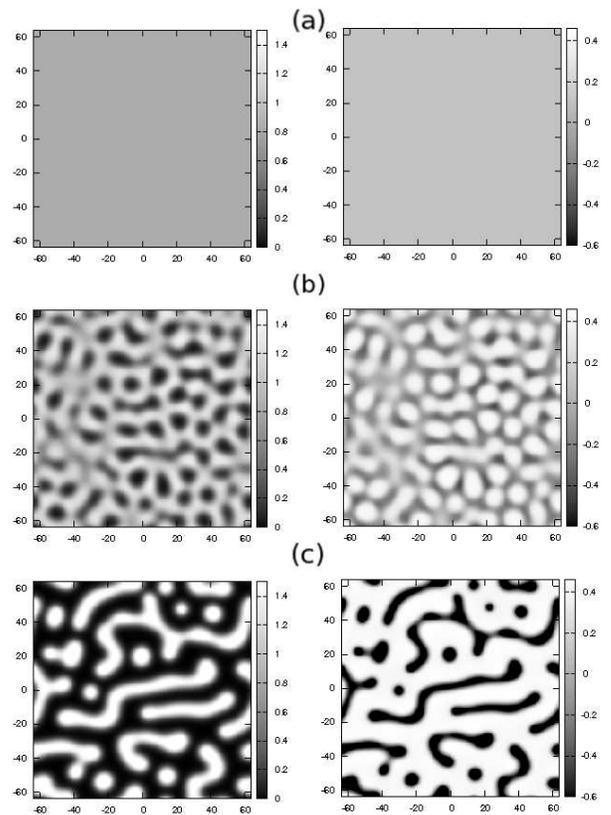}
  \caption{Time evolution of the order parameter
  $\psi$ (left) and salinity $\rho$ (right) versus spatial 
  coordinates for $\tau=10, 150, 500$ (from above to below)
  with the initial random distribution $\psi(\tau=0)=0.9$
  and $\rho(\tau=0)=0.1 \pm 0.001 N(0,1)$. The parameters
  are $\alpha_3 = 0.9$, $\alpha_1' = 0.1$, and $D = 0.5$.}
\label{Num2D}
\end{figure}

For the one-dimensional case, we plot snapshots of the time evolution of the order 
parameter $\psi$ and the salinity $\rho$ in Fig. \ref{Num1D}. The evolution of the order 
parameter $\psi$ and the salinity $\rho$ in two dimensions is shown in Fig. \ref{Num2D}. 
The quantities $\psi$ and $\rho$ are opposite in phase to each other. Regions of high 
salinity correspond to the water phase and regions of low salinity to ice domains. 
During the freezing process we observe a desalinization of ice and an increase of the 
salinity in the liquid phase.
In the numerical solution a small initial random perturbation is applied to the 
salinity $\rho(\tau=0)$ normally distributed having zero mean and variance one. 
In Fig. \ref{Num1D} we see that one single mode develops given by the wave 
number $\kappa_c$. Similar to the one-dimensional case, we see the formation of one 
dominant wavelength also in two dimensions (Fig. \ref{Num2D}). In the next chapter, we compare 
the model result for the pattern size to the experimental quantities.

\section{Comparison with the experiment}

\subsection{Determination of parameters from water properties}
\subsubsection{Temperature-dependent representation}

The thermodynamic function of state selected for the phase field simulation depends on 
the definition of the problem. Entropy is appropriate for a system of constant energy and 
volume, the Helmholtz free energy is the proper choice for an isothermal system at constant
volume, and the Gibbs free energy when the temperature and pressure are kept constant.
The phase field equations of our study are in terms of the Landau-Ginzburg free energy 
functional (\ref{free_energy}). The Landau-Ginzburg free energy functional can be thought of 
as a Helmholtz free energy or as a Gibbs free energy. We consider the Gibbs free energy as 
the short-time structures are usually formed under constant pressure, and the experimental 
data of water properties are based on isothermal-isobar conditions. Now we are going to 
discuss the two crucial parameter $\alpha_1$ and $\alpha_3$ in terms of their temperature
dependence. In this way we will be able to extract the physical meaning of the different
instability regions and link the chosen values of the parameters to the supercooling 
and superheating as well as freezing point temperature of water. The idea consists in 
identifying the upper borderline of stable structure formation (\ref{16}) with the
freezing point temperature since this is the line where structure, i.e. ice
formation is possible at all.  In the same manner we identify the borderline
of metastable structure (\ref{inst}) with the superheating temperature.
 
\begin{figure}[h]
\includegraphics[width=8cm,angle=0]{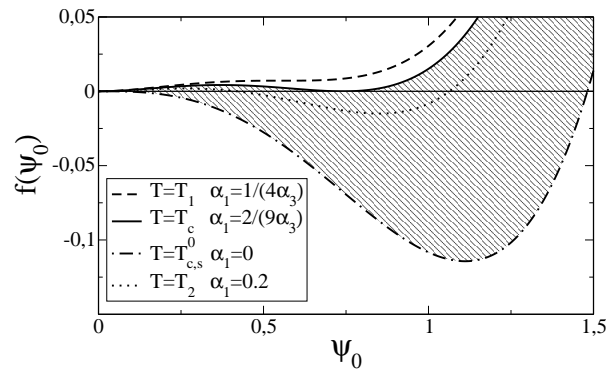}
\caption{Representation of supercooling region for freezing of
  hexagonal ice $T^0_{c,s}<T_c$ (shaded area) and the superheating 
region  $T_c<T<T_1$.\label{free_energy_temp}}
\end{figure}

The freezing point temperature depends on the salinity. 
The solidification of seawater is approximated as that of a dilute aqueous NaCl 
solution, since Na$^+$ and Cl$^-$ are the dominant ions in seawater. Hence, the mean salinity 
in seawater of $35$ g/kg corresponds to 1 NaCl molecule per 100 H$_2$O-molecules, 
i.e. 1 Na$^+$ and 1 Cl$^-$ per 100 H$_2$O-molecules in solution after the dissociation or a ratio 
of $x=({n_{\rm Na^+}+n_{\rm Cl^-}})/{n_{\rm H_2O}}=1/50$. 
Using the Clausius-Clapeyron (cc) relation
\begin{equation}
 \Delta T_{cc} = -\frac{xRT^2}{\Delta H}\, , 
\label{size4}
\end{equation}
where $\Delta H$ = 6\ $\frac{\rm kJ}{\rm mol}$ is the latent heat of the phase transition
from water to ice, $R$=8.314\ $\frac{\rm J}{\rm mol\, K}$ is the universal gas constant and
$T$=273\ K, we obtain a freezing point depression from 0$^\circ$C to -2$^\circ$C in 
agreement with the natural value of $\Delta T$ = -1.9\ K. 
The temperature and salinity dependence of the phase transition is described by the 
coefficient of the quadratic term in $\psi$ of the Landau-Ginzburg functional 
\begin{equation}
\alpha_1(T)=\alpha_1'(T) + \rho_0\, . 
\label{size1}
\end{equation}
Because of the requirement that $\alpha_1'(T_c^0) = 0$ at the lower limit of the
supercooling region of fresh water, $T_c^0$ = 233.15\ K (\cite{Nev,Dor}), 
the parameter $\alpha_1'(T)$ is approximated by 
\begin{equation}
\alpha_1'(T) = \tilde{\alpha}_1(T-T_c^0)\, , 
\label{size2}
\end{equation}
where $\tilde{\alpha}_1$ is a temperature-independent coefficient. The temperature and 
salinity dependence of the quartic term of the dimensionless Landau-Ginzburg functional 
is supposed to be so weak near the phase transition that it can be well approximated as 
a positive constant, namely, $\alpha_3$. 
Using the equations (\ref{size1}), (\ref{size2}) and (\ref{freezp}), we obtain for the 
freezing point depression in the framework of the Landau-Ginzburg theory the expression
\begin{equation}
 \Delta T = -\frac{\rho_0}{\tilde{\alpha}_1} = 
            -\frac{b_1}{\tilde{a}_1}\frac{a_2^2}{h^2}\rho_0\, .
\label{size3}
\end{equation}
Introducing the salinity-dependent supercooling temperature
\begin{equation}
T^0_{c,s}=T_c^0-|\Delta T|
\end{equation}
the freezing parameter $\alpha_1$ depends on the temperature according to
\begin{equation}
\alpha_1(T) = \rho_0\frac{T-T^0_{c,s}}{|\Delta T|}\, .
\label{aa0}
\end{equation}
Besides the supercooling temperature we have two further specific values of the 
freezing parameter $\alpha_1$ corresponding to two additional temperatures
\begin{eqnarray}
\alpha_1(T_1)&=&\rho_0\frac{T_1-T^0_{c,s}}{|\Delta T|}=\frac{1}{4\alpha_3} 
\label{aa1}\\
\alpha_1(T_c)&=&\rho_0\frac{T_c-T^0_{c,s}}{|\Delta T|}=\frac{2}{9\alpha_3} 
\label{aa2} 
\end{eqnarray}
defining the temperatures $T_1$ and $T_c$. The shaded area in
Fig. \ref{free_energy_temp} describes the supercooling region between $T_c$ and
$T^0_{c,s}$. Above this area we find the superheating region for $T_c<T<T_1$.

\begin{figure}[h]
\includegraphics[width=8cm,angle=0]{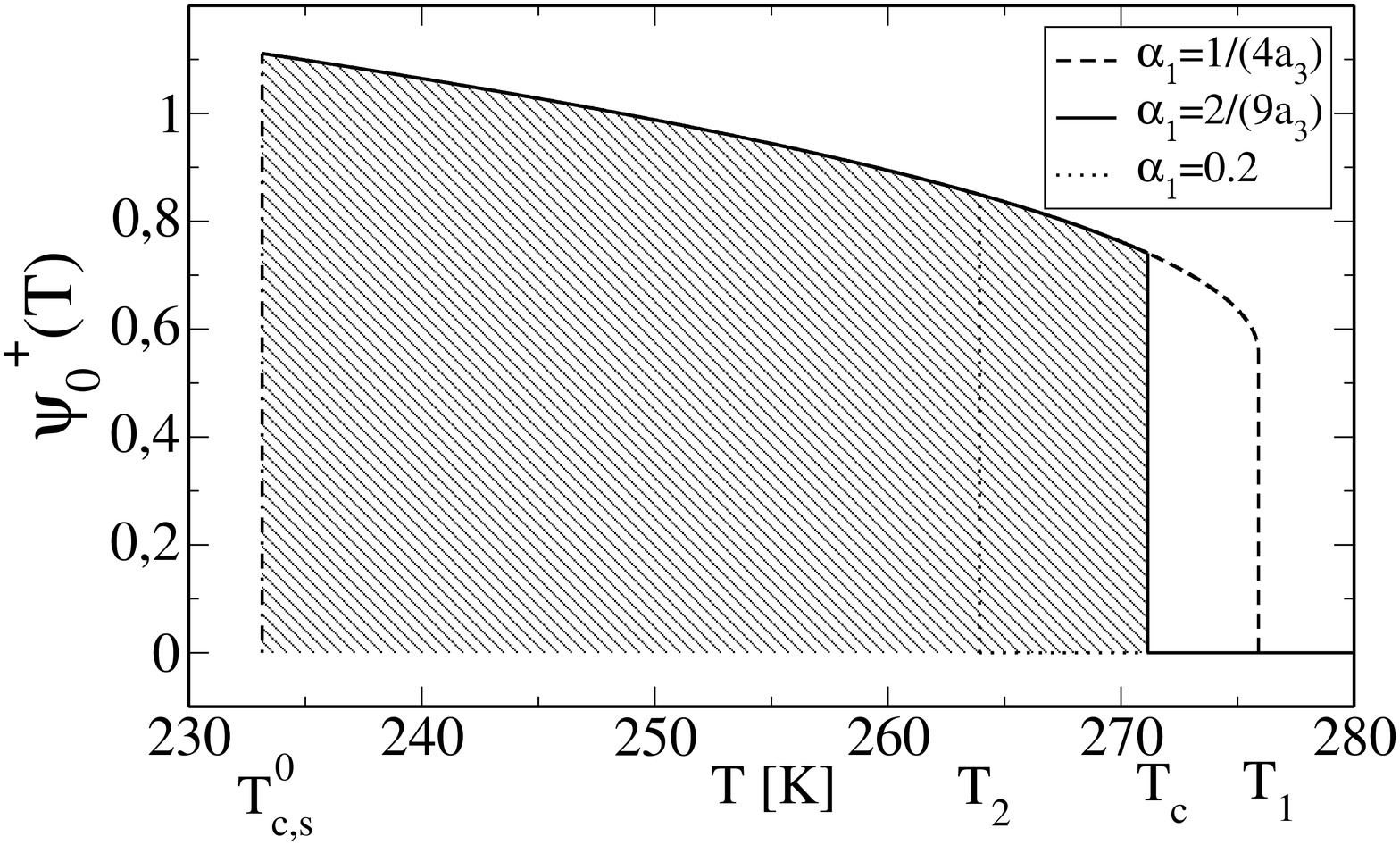}
\caption{Temperature-dependent representation of the supercooling
  region ($T<T_c$) and the superheating region ($T>T_c$). \label{psi_temp}}
\end{figure}

For the relative minimum for the order parameter $\psi_0^+$ in (\ref{psi0p})
we obtain with respect to (\ref{aa0}) and (\ref{aa2}) the
temperature dependence
\begin{equation}
\psi_{0}^+(T)=\frac{3}{4}\psi_{0,c}\left(1+
\sqrt{1-\frac{8}{9}\frac{T-T^0_{c,s}}{T_c-T^0_{c,s}}}\right)\, .
 \label{aa4}
\end{equation}
where the position $\psi_{0,c}=\psi_0(T=T_c)=\frac{2}{3\alpha_3}$ of the coexistence 
line corresponds to the freezing point temperature $T_c$. Analogously to 
Fig. \ref{FreeEnergy} we plot in Fig. \ref{psi_temp} the relative minimum $\psi_0^+$ 
of the free energy but now versus temperature. 

Alternatively we can express the fixed point by 
the superheating temperature (\ref{aa1}) as well since
\begin{equation}
4 \alpha_1(T)\alpha_3=
\frac{8}{9} \frac{T-T^0_{c,s}}{T_c-T^0_{c,s}}=\frac{T-T^0_{c,s}}{T_1-T^0_{c,s}}.
\end{equation}
This leads to the relation between the supercooling temperature $T^0_{c,s}$, the
freezing point temperature $T_c$ and the superheating temperature $T_1$
\begin{equation}
T_1=\frac{9}{8}T_c - \frac{1}{8}T^0_{c,s}\, .
\label{aa5}
\end{equation}
In contrast to the supercooling of the water it is difficult to superheat an ice crystal.
After a superheating of more than 5$^\circ$C homogeneous nucleation occurs in 
the metastable state \cite{Baum}. 
For fresh water ($T_c=T_0$ = 273.15\ K and $T^0_{c,s}=T_c^0$ = 233.15\ K) if follows 
from equation (\ref{aa5}) $T_1$ =278.11\ K (4.96$^\circ$C) as 
the upper limit of superheating in agreement with the experiment \cite{Baum}.

Using (\ref{aa0}) and (\ref{aa2}) we can explain the values of $\alpha_1$ and $\alpha_3$ 
chosen for the numerical solution of the model equations in terms of water properties. 
We chose the structure parameter $\alpha_3=0.9$ such that for seawater of 
salinity 35\ g/kg ($\rho_0$ = 0.6 mol NaCl/53 mol H$_2$O = 0.0113) the freezing point 
temperature is -1.9$^\circ$C ($T_c$ = 271.25\ K). 
While dominated by Na$^+$ and Cl$^-$, seawater contains several other ions that influence its 
phase behaviour. Seawater is a multicomponent system that involves the crystallisation of some
salts close to their respective eutectic concentrations. For temperatures $0 > T > -8^\circ$C, 
the influence of precipitating Ca$^{2+}$ ions (in form of CaCO$_3$6H$_2$O) is almost negligible, 
but the onset of heavy precipitation of Na$^+$ ions (in form of mirabilite: Na$_2$SO$_4$10H$_2$O)
below -8$^\circ$C makes the theoretical description of the phase relationships in seawater more
complicated. The precipitation of salt is not described by the presented model. Therefore, the
model can be used to predict solidification structures only for the high temperature regime down 
to $\approx -8^\circ$C. The size of solidification structures depends on the supercooling 
relative to the freezing point $T_c$. The higher the supercooling, the more rapidly the water
freezes and the smaller are the structures. Hence, discussing the time evolution of the salinity
distribution and ice structure it is useful to perform the numerical simulation for high
supercooling (but not below the precipitation temperature of mirabilite). Our choice of the
freezing parameter $\alpha_1=0.2$ represents a temperature $T_2$ = -8.2$^\circ$C in 
agreement with a frequently cited transition temperature of mirabilite. 
This along with (\ref{aa5}) shows that with a structure parameter $\alpha_3=0.9$ and a 
freezing parameter $\alpha_1=0.2$ one has a realistic description of seawater at a 
temperature close to -8$^\circ$C in terms of superheating and supercooling of pure water. 
For this case the time evolution of the 2D patterns will be simulated with reasonable 
computing time using a small $128 \times 128$ grid. The morphology of the phase field 
structure as $T$ approaches $T_c$ is likely not altered, in particular 
because for $T_c > T > T_2$ a structural phase transition does not occur. However, the size of 
the structure ranges from $\mu$m to mm scale as discussed in the next section.

\subsubsection{Diffusivity}

The parameters $\alpha_1$ and $\alpha_3$ define the local portion of the free 
energy (\ref{fzdiff}), which describes a system with uniform order parameter and salinity. 
The spatial inhomogeneity of the system is described by the third parameter of the 
model $D=D_{\rm ice}/D_{\rm salt}$. Spatial fluctuations of the order parameter
are controlled by the diffusivity-type constant $D_{\rm ice}={d_c^2}/{6 \tau}$, 
with the correlation length $d_c$ and the relaxation time $\tau$ near the freezing point. 
We estimate the correlation length from the equilibrium radius $r_{i/w}$ of a spherical 
ice embryo formed from supercooled water near the freezing point. More recent 
theoretical simulations of ice nucleation involving the curvature effect on ice surface tension 
predict $r_{i/w} = 10^{-7}$\ m (Fig. 5 in \cite{Bogd}). The classical nucleation theory has not 
been tested for ice crystal nucleation since there are no independent measurements of the ice
surface tension. Instead, the validity of the theory has been assumed and the measured 
freezing point temperature has been used to obtain the value of the surface 
tension (\cite{Bogd} and ref. therein). The relaxation rate ${1}/{\tau}$ is proportional 
to reorientations of the H$_2$O-molecules per second, ${1}/{\tau_d(T)}$, which at the freezing 
point of fresh water assumes the value ${1}/{\tau_d(T_0)} = 0.5\times 10^5$\ s$^{-1}$ \cite{Eis}. 
With $d_c =2\times r_{i/w}$ and $\tau = \tau_d(T_0)$ it 
follows $D_{\rm ice} = 0.33\times 10^{-5}$\ cm$^2$/s.  
The diffusion constant of salt $D_{\rm salt}$ we estimate from the counter-diffusion coefficient 
of the dominating salt in seawater NaCl
\begin{equation}
D_{\rm NaCl}=
\frac{\left(|Z_{\rm Na^+}|+|Z_{\rm Cl^-}|\right)D_{\rm Na^+}D_{\rm Cl^-}}
{|Z_{\rm Na^+}|D_{\rm Na^+}+|Z_{\rm Cl^-}|D_{\rm Cl^-}}\, ,
\label{size7}
\end{equation}
where $D_{\rm Na^+}$ and $D_{\rm Cl^-}$ are the tracer-diffusion coefficients 
of Na$^+$ and Cl$^-$, respectively. $Z_{\rm Na^+}$ and $Z_{\rm Cl^-}$ are the charges 
of the respective ions. Using the tracer-diffusion coefficients reported in \cite{Li74} 
for 0$^\circ$C at infinite dilution, $D_{\rm Na^+}=6.27\times 10^{-6}$\ cm$^2$/s and 
$D_{\rm Cl^-}=10.1\times 10^{-6}$\ cm$^2$/s, it follows
$D_{\rm NaCl}=0.77\times 10^{-5}$\ cm$^2$/s. 
Generally, the solute diffusion coefficient of aqueous solutions displays a weak 
salinity dependence.
The diffusion coefficient in seawater is only about 8 per cent smaller than in 
water \cite{Li74}, i.\ e.\ $D_{\rm salt,0^\circ C}=0.71\times 10^{-5}$\ cm$^2$/s.
Most studies of aqueous NaCl have been restricted to temperatures above 0$^\circ$C,
and for $D_{\rm salt}$ only a few datasets extend down to 0$^\circ$C.
The required extrapolation to subzero temperatures can be found in \cite{Ma07}.
At the freezing point temperature of seawater, -1.9$^\circ$C, the study 
in \cite{Ma07} predicts $D_{\rm salt,-1.9^\circ C}=0.62\times 10^{-5}$\ cm$^2$/s.
The ratio $D_{\rm ice}/D_{\rm salt}=0.53$ obtained for the computed $D_{\rm salt}$ 
at -1.9$^\circ$C may be compared to $D_{\rm ice}/D_{\rm salt}=0.47$ based on the 
measured $D_{\rm salt}$ at 0$^\circ$C.
Hence we estimate for the model parameter $D=D_{\rm ice}/D_{\rm salt}=0.5$.
%
%

\begin{figure*}[tb]
\centerline{\includegraphics[width=.65\textwidth,angle=0]{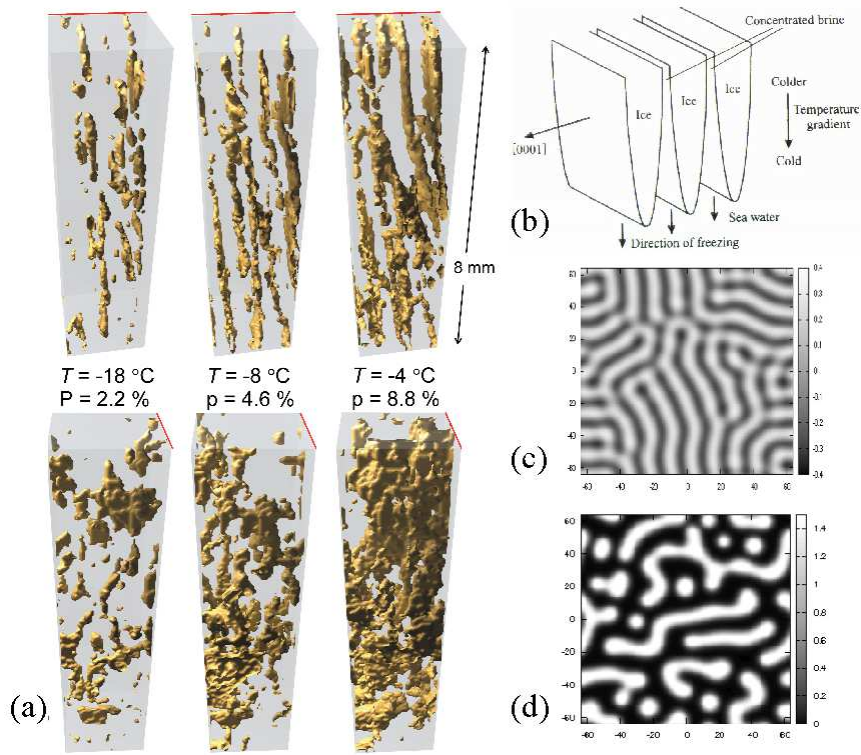}}
\caption{\label{comp} 
%
%
(a) Imaging brine pore space with X-ray computed tomography (image from \cite{Pring09}). 
The upper images shows the view approximately along the brine layers. 
The view across the brine layers (of the same volumes rotated about the vertical 
axis by $90^\circ$) is shown in the bottom images. 
(b) Sea ice crystal composed of parallel plates of pure ice with concentrated brine 
in layers between them (image from \cite{Pet}). 
(c) Turing structure for $\tau=400$ \cite{KMG09}, 
(d) phase field structure for $\tau=500$ from figure \ref{Num2D}.
}
\end{figure*}

\subsection{Brine layer microstructures}

As the two-dimensional interstitial liquid freezes, small ice 
bridges form between adjacent ice platelets, trapping inclusions of brine.
Now that we have determined the three model parameters from the properties of water
we can proceed and simulate the brine entrapment using our phase-field model. 
In figure \ref{Num2D} we show the time evolution of the salinity distribution 
and ice structure in the intercrystalline brine layer. We observe that structures 
found here are relative stable for a long time scale such that we assume them to be frozen.
The final result of the structure is plotted in Fig. \ref{comp}(d).

{\sc Pringle} {\it et al.} \cite{Pring09} used sea ice single crystals, the building 
blocks of polycrystalline sea ice, to image their complex pore 
space with X-ray computed tomography. The structure of the brine 
pore space is seen in isosurface plots in Fig. \ref{comp}(a). The 
upper images clearly show near-parallel intracrystalline brine layers.
However, the view across the layers (bottom images) show brine 
layer textures much more complicated than suggested by the 
simple model of parallel ice lamellae and parallel brine sheets
illustrated in Fig. \ref{comp}(b). 
Images at -18$^\circ$C, -8$^\circ$C and -4$^\circ$C show 
a thermal evolution of the brine pore space where the 
porosity $p$ changes from $2.2$ to $4.6$ to $8.8\%$. The connectivity 
increases with porosity as the pore space changes from isolated 
brine inclusions at $p = 2.2\%$ to extended, near-parallel layers 
at $p = 8.8\%$. {\sc Pringle} {\it et al.} \cite{Pring09} characterized the thermal 
evolution of the brine pore space with percolation theory and demonstrated a 
connectivity threshold at a critical volume fraction $p_c = 4.6\%$. 
Below $p_c$ there are no percolating pathways spanning a sample, i.e.
the brine is trapped within the intracrystalline brine layers. 
In Fig. \ref{comp}(c) we have chosen the best fit of the Turing-model 
\cite{KMG09} to the brine layer structure one observes very close to $p_c$. 
If we compare the simulation according to the here presented phase field 
model in Fig. \ref{comp} (d) the brine entrapment is better described by 
the short-time phase-field structures than by the Turing image.
 
Please note that the three parameters of the Turing model had been adjusted 
to fit the structure as best as possible. Here with the 
phase-field model we have chosen parameters according to the thermodynamic 
properties of water and have obtained the structure with these parameters. 
That the physical justification of the parameters by other properties of water 
leads here to a better description of the brine layer texture we 
consider as a demonstration of the importance of mass conservation invoked in 
the present model. 

Now we are going to compare the size of the obtained structure with the size of pure 
sea ice platelets, separating regions of concentrated seawater. Sea ice platelet spacing depends 
on the growth velocity of ice crystals in the basal plane \cite{We92}, which is given as a 
function of salinity and supercooling $\Delta T_{sup}$(relative to the freezing point $T_c$). 
The growth velocity of ice crystals is described by the empirical 
law $V= \gamma\times 1.87\times 10^{-2}\Delta T_{sup}^{2.09}$, with $V$ given in cm/s. 
Crystal growth in pure water is obtained for $\gamma =1$ \cite{TirGi}. Ionic solutes do not 
considerably affect the growth kinetics, though the actual growth velocity at a 
given $\Delta T_{sup}$ will be changed by the factor $\gamma$. For NaCl 
solutions  $> 10$\ g/kg, $\gamma$ falls monotonically with increasing solute concentration 
\cite{VlaBa}, a solution of 35\ g/kg may decrease $V$ by 50\%. During phase separation, the 
water adjacent to the growing ice crystal becomes enriched in the rejected salt. 
The higher interfacial salt concentrations tend to depress the crystallization velocity 
more severely. The interface concentration is limited by the eutectic value 
of $\approx$230\ g/kg. A NaCl solution of eutectic composition reduces the growth velocity 
by up to 86\%. With $\gamma = 0.14$ and the chosen 
supercooling $\Delta T_{sup} = T_c -T_2$ = 6.3\ K we 
obtain $V= \gamma\times 1.87\times 10^{-2}\Delta T_{sup}^{2.09}=1.2\times 10^{-1}$\ cm/s. 
Using the result for $V$, the size of sea ice platelets can be obtained from the 
morphological stability theory. For $V\approx0.1$\ cm/s, the morphological stability 
analysis predicts a wave length of most rapidly growing perturbations 
of $\lambda_{max}\approx1$\ $\mu$m (see Fig. 2 in \cite{We92}). 

Now we use equation (\ref{size3}) to calculate the critical domain size of 
the phase field structure as a function of the freezing point depression. The critical domain 
size is determined by the fastest-growing wave-vector $\kappa_c$ (\ref{kcrit}). 
From the relation between the dimensionless wavenumber $\kappa$, the dimensional wavenumber $k$ 
and equation (\ref{size3}) we get the minimum size of the structure
\begin{equation}
\lambda_c = \frac{2 \pi}{k_c}=\frac{2 \pi}{\kappa_c}\sqrt{\frac{M}{\Gamma}}\frac{h}{a_2}
    = \frac{2 \pi}{\kappa_c}\sqrt{\frac{D_{\rm salt}\rho_0}{\Gamma\tilde{a}_1|\Delta T|}}\, ,
\label{size6}
\end{equation}
with $2\pi/\kappa_c = 13.81$ and the parameters specified in the previous 
section:  $D_{\rm salt}=0.71\times 10^{-5}$\ cm$^2$/s, $\rho_0=0.0113$, and $|\Delta T|=1.9$. 
The parameter left to be determined is $\Gamma \tilde{a}_1$. This parameter we derive from the 
temperature dependence of the order parameter kinetics (\ref{lg_1}), which is described by the 
rate constant $\Gamma a_1'= \Gamma a_1'(T)$ for the change of the dimensional order 
parameter $u$. The rate $\Gamma a_1'(T)$ is proportional to reorientations of 
the H$_2$O-molecules per second, ${1}/{\tau_d(T)}$, which at the freezing point of fresh 
water, $T_0$ = 273.15\ K, assumes the 
value ${1}/{\tau_d(T_0)} = 0.5\times 10^5$\ s$^{-1}$ \cite{Eis}. Because of 
the requirement that $a_1'(T_c^0) = 0$ at the lower limit of the supercooling 
region, $T_c^0$ = 233.15\ K, the rate $\Gamma a_1'(T)$ is approximated by 
\begin{equation}
\Gamma a_1'(T) = \Gamma \tilde{a}_1(T-T_c^0) \sim \frac{1}{\tau_d(T)}\, .
\label{size8}
\end{equation} 
The temperature-independent coefficient $\Gamma \tilde{a}_1$ we estimate from 
$\Gamma a_1'(T_0)=\frac{1}{\tau_d(T_0)}$, i. e.
we assume $\Gamma \tilde{a}_1 = 1/[(T_0-T_c^0)\tau_d(T_0)] = 1250$\ K$^{-1}$\ s$^{-1}$. 
Our choice of the freezing parameter $\alpha_1=0.2$ represents a 
supercooling $\Delta T_{sup} = 6.3$\ K. For this growth condition we obtain from 
equation (\ref{size6}) a critical domain size $\lambda_c = 0.8$\ $\mu$m in agreement with 
the sea ice platelet spacing $\lambda_{max}\approx1$\ $\mu$m obtained from morphological 
stability analysis \cite{We92}.
Please note that our choice of 6.3\ K supercooling does not represent the
natural conditions observed in the oceans. Supercoolings are commonly believed 
to be small ($<0.1$\ K) owing to the abundant presence of impurities upon which 
nucleation of ice crystals occurs via heterogeneous mechanisms \cite{PetEi10}. 
Heterogeneous nucleation processes are responsible for virtually all of the 
ice on earth. 

In nature, fresh water is never totally pure. While supercooling in 
pure water has been observed down to -40$^\circ$C, the presence of small impurities 
commonly initiate freezing at temperatures between -15$^\circ$C 
and -20$^\circ$C \cite{Hag81}. The observed sea ice plate spacings 
are $2-3$ orders of magnitude larger than 1\ $\mu$m. The thickness of ice lamellae 
between brine layers reported in \cite{Pring09} was in the range 
of 200-500\ $\mu$m (see Fig 1a). Brine inclusions described 
by {\sc Weissenberger} \cite{Wei} had scales from below $3$ to 1000\ $\mu$m, where the 
average dimensions were typically 200\ $\mu$m. The size of the phase-field structures 
for natural conditions are described by the upper limit of the instability region shown 
in Fig. \ref{InstabReg}. With a structure parameter $\alpha_3=1.99$ and a freezing 
parameter $\alpha_1=0.111482$ one has a realistic description of seawater 
at 0.032\ K supercooling and a lower limit of the supercooling region of 
fresh water at -18.78$^\circ$C. For this growth consition we obtain from 
equation (\ref{kcrit}) the dimensionless structure size $2\pi/\kappa_c = 4975.25$.
Using equation (\ref{size6}) we obtain a critical domain size $\lambda_c$ = 198\ $\mu$m 
in agreement with the observed values.

\section{Conclusion}

A phase field model for brine entrapment in sea ice is developed which describes the 
time evolution of the micro-scale inclusions in the absence of spatial temperature 
gradients in ice or water. The coupled evolution equations are derived from a generating
functional in such a way that the mass conservation law for salt is preserved. The 
resulting equations are different from the earlier used Turing model which does not 
include conservation of salt. Though the same number of three parameters are used the phase 
field model allows to determine them by physical properties and lead to a better agreement 
with experimental brine layer textures.

The linear stability analysis provides the phase diagram in terms of two parameters 
indicating the region where spatial structures can be formed due to the instability of the 
uniform ordered phase. The region of instability is determined exclusively by the 
freezing parameter and the structure parameter and not by the diffusivity as it was the case 
in the Turing model. This allows to link the freezing and structure parameter to 
thermodynamic properties of water like superheating, supercooling and freezing temperature. 

With the help of the parameters determined by the properties of water we solve the 
time-dependent coupled evolution equations and determine the micro-scale brine network. 
Since, the brine tends to be located in vertically oriented sheet-like inclusions, we 
performed the numerical simulations in two dimensions. A good agreement with the 
experimental brine layer texture is achieved. Therefore we believe that the simulation of 
the short-time frozen microstructures led to a realistic description of the early phase of 
brine entrapment. Brine inclusions that evolve in the long-term show length scales from 
sub-millimeter brine layers to meter long brine channels. The description of the brine 
channel formation requires simulations on large scales in three-dimensional domains including 
a detailed description of the thermal evolution of brine inclusions. The quite demanding 
large scale simulations will be the subject of future research.

\acknowledgments

This work was supported by DFG-priority program SFB 1158. The financial support by the Brazilian
Ministry of Science and Technology is acknowledged.


\end{document}